\documentclass[letterpaper,titlepage,11pt]{article}
\usepackage[colorlinks=true,urlcolor=blue,citecolor=magenta]{hyperref}
\usepackage{amssymb,amsmath,amsfonts}
\usepackage{epsfig}
\usepackage{epsfig}
\usepackage{graphicx}
\usepackage{epstopdf}
\def\clock{{\count0=\time
           \divide\count0 60
           \ifnum\count0<10 0\fi\the\count0
           \multiply\count0 -60 \advance\count0 \time
           :\ifnum\count0<10 0\fi \the\count0
         }}
\newcommand{\timestamp}{{\small\vbox{\hbox{\tt\jobname.tex}
\hbox{\the\day/\the\month/\the\year, \clock}}}}

\newcommand{\CB}{\mathcal{B}}

\newcommand{\CD}{\mathcal{D}}

\newcommand{\CL}{\mathcal{L}}

\newcommand{\CO}{\mathcal{O}}

\newcommand{\CR}{\mathcal{R}}

\newcommand{\CW}{\mathcal{W}}

\newcommand{\nn}{\nonumber}

\newcommand{\spa}{\ , \ \ }

\newcommand{\be}{\begin{eqnarray}}
\newcommand{\ee}{\end{eqnarray}}
\newcommand{\beq}{\begin{eqnarray}}
\newcommand{\eeq}{\end{eqnarray}}

\newcommand{\beqa}{\begin{eqnarray}}
\newcommand{\eeqa}{\end{eqnarray}}

\newcommand{\ds}{\displaystyle}
\newcommand{\bk}{\mathbf{k}}

\let\oldsqrt\sqrt
\def\sqrt{\mathpalette\DHLhksqrt}
\def\DHLhksqrt#1#2{%
\setbox0=\hbox{$#1\oldsqrt{#2\,}$}\dimen0=\ht0
\advance\dimen0-0.2\ht0
\setbox2=\hbox{\vrule height\ht0 depth -\dimen0}%
{\box0\lower0.4pt\box2}}

\newcommand{\ads}{\mbox{AdS}}

\numberwithin{equation}{section}

\begin{document}

\begin{titlepage}
\rightline{\vbox{\small\hbox{\tt NORDITA-2012-53} }}
\rightline{\vbox{\small\hbox{\tt NSF-KITP-12-128} }}
 \vskip 1.8 cm

\centerline{\Huge \bf Thermal Giant Gravitons}

\vskip 1.5cm

\centerline{\large {\bf Jay Armas$\,^{1}$},  {\bf Troels Harmark$\,^{2}$}, {\bf Niels A. Obers$\,^{1}$},}
\vskip 0.2cm \centerline{\large   {\bf Marta Orselli$\,^{1}$} and
{\bf Andreas Vigand  Pedersen$\,^{1}$} }

\vskip 1.0cm

\begin{center}
\sl $^1$ The Niels Bohr Institute, Copenhagen University  \\
\sl  Blegdamsvej 17, DK-2100 Copenhagen \O , Denmark
\vskip 0.4cm
\sl $^2$ NORDITA\\
Roslagstullsbacken 23,
SE-106 91 Stockholm,
Sweden 
\end{center}
\vskip 0.6cm

\centerline{\small\tt jay@nbi.dk, harmark@nordita.org, obers@nbi.dk}
\centerline{\small\tt orselli@nbi.dk, vigand@nbi.dk}

\vskip 1.3cm \centerline{\bf Abstract} \vskip 0.2cm \noindent
We study the giant graviton solution as the $\ads_5\times S^5$
background is heated up to finite temperature.  
The analysis employs the thermal brane probe technique based on the blackfold approach.
We focus mainly on the thermal giant graviton corresponding to a thermal D3-brane
probe wrapped on an $S^3$ moving on the $S^5$ of the background at finite
temperature. We find several interesting new effects, including that
the thermal giant graviton has a minimal possible value for the
angular momentum and correspondingly also a minimal possible radius of
the $S^3$. We compute the free energy of the thermal giant graviton
in the low temperature regime, which potentially could be compared to that of a thermal state on the
gauge theory side. Moreover, we analyze the space of solutions and
stability of the thermal giant graviton and find that, in parallel with the extremal case, there are two
available solutions for a given temperature and angular momentum, one
stable and one unstable. In order to write down the equations of
motion, action and conserved charges for the thermal giant graviton we
present a slight generalization of the blackfold formalism for charged
black branes. Finally, we also briefly consider the thermal giant
graviton moving in the $\ads_5$ part.

\end{titlepage}


\tableofcontents

\section{Introduction}

Recently, a new method to study thermal brane probes in string/M-theory has been explored. This method consists in using the blackfold approach \cite{Emparan:2009cs,Emparan:2009at}%
\footnote{See also Ref. \cite{Emparan:2007wm} which discussed the first application to neutral black rings in asymptotically flat space. Reviews include \cite{Emparan:2009zz} and  \cite{Camps:2012hw}  gives a more general derivation of the blackfold effective theory.} in the context of 
 string/M-theory branes \cite{Emparan:2011hg,Caldarelli:2010xz,Grignani:2010xm}. 
In the blackfold approach one can describe the dynamics of a black brane wrapped on a submanifold of the background space-time in the probe approximation where the black brane is much thinner than the length scale of the submanifold. This method has been applied to the thermalized version of the BIon system for the D3-brane \cite{Grignani:2010xm,Grignani:2011mr}, 
the gravity dual of the rectangular Wilson loop as described by an F-string ending on the boundary of $\ads_5\times S^5$ \cite{Grignani:2012iw}, the M2-M5 version of the BIon system \cite{Niarchos:2012pn}, and has been used to find a number of new black holes in String/M-theory \cite{Emparan:2011hg,Caldarelli:2010xz}.%
\footnote{Large classes of new neutral black objects in asymptotically flat and $(A)dS$ backgrounds were
found in  \cite{Emparan:2009vd} and \cite{Caldarelli:2008pz} respectively.}

In this paper we apply the method to giant gravitons \cite{McGreevy:2000cw,Grisaru:2000zn,Hashimoto:2000zp}.
 The archetypical case of a giant graviton is that of a D3-brane wrapped on a three-sphere and with the center of mass moving along the equator of the five-sphere in the $\ads_5\times S^5$ background. This corresponds to a blown up version of a point particle graviton and is described using the Dirac-Born-Infeld (DBI) action for the extremal D3-brane.  Our goal is to analyze what happens to the giant graviton as one heats up the $\ads_5\times S^5$ background to non-zero temperature, requiring the brane probe to thermalize with the background. We dub the resulting brane 
probe a {\sl thermal giant graviton}. 

In the dual gauge theory description of $\ads_5\times S^5$ the giant graviton moving along the equator of $S^5$ with angular momentum $J$ is dual to a gauge theory multi-trace operator $\CO_{\rm gg}$ with R-charge $J$ and conformal dimension $\Delta = J$.  
Heating up the $\ads_5\times S^5$ background and with it the giant graviton brane probe corresponds then on the gauge theory side to the thermal state that results from the ensemble of operators that are fluctuations around $\CO_{\rm gg}$. This is true up to the temperature $T_{\rm HP}$ of the Hawking-Page transition where an $\ads$ black hole is formed. Thus, having a description of a thermal giant graviton will provide important insight into the strong coupling behavior of the gauge theory side at finite temperature.

We find in this paper the following features for the thermal giant graviton corresponding to a black D3-brane probe wrapped on a three-sphere with center of mass moving along the equator of $S^5$ in the $\ads_5\times S^5$ background. Firstly, we find for a given temperature that the family of solutions has $J \geq J_{\rm min} > 0$. Instead for the extremal giant graviton one can take the limit $J \rightarrow 0$ thus connecting to the point particle graviton. The thermal giant graviton, however, is forced to be blown up to a finite-size three-sphere. This is analogous to what happens in the thermal BIon case where the throat of the brane - made of a D3-brane wrapped on an $S^2$ - has a minimal possible radius \cite{Grignani:2010xm,Grignani:2011mr}. Secondly, we find a maximal temperature $T_{\rm max}$ which provides the scale of the temperature of the solution. However, the probe approximation gives $T_{\rm max} \gg T_{\rm HP}$, hence requiring $T \leq T_{\rm HP}$ means that we have small temperature for the thermal giant graviton $T / T_{\rm max} \ll 1$. This is analogous to  the case of the thermal rectangular Wilson loop of \cite{Grignani:2012iw}. Thirdly, for the free energy we find the following expansion for $T / T_{\rm max} \ll 1$
\begin{equation} 
\label{Fresult}
F (T,J) = \frac{J}{L} - \frac{\pi^4}{4}  N_{\rm D3}^2 L^3 T^4   +{\cal{O}} (T^8)  \ . 
\end{equation}
This is a prediction at strong coupling which can potentially be compared to what one can compute on the gauge theory side. Finally, we note that the phase structure of the family of solutions at finite temperature is similar to the extremal case in that for a given temperature and for $J_{\rm min} < J < J_{\rm max}$ one has two available solutions, one stable and one unstable. This is different than for the thermal BIon and the thermal rectangular Wilson loop cases in that there the number of available phases could become higher or lower when turning on the temperature \cite{Grignani:2010xm,Grignani:2011mr,Grignani:2012iw}.

Since we are employing the blackfold approach for studying the thermal giant graviton our black D3-brane probe consists of $N_{\rm D3}$ coincident black D3-branes in the supergravity approximation with $N_{\rm D3} \gg 1$ and $N \ll \lambda N_{\rm D3}$. At the same time the probe approximation requires $N_{\rm D3} \ll N$.

In order to describe the thermal giant graviton using the blackfold approach we need to generalize the blackfold approach slightly. Firstly, we write down the blackfold equations of motion for charged branes in backgrounds with fluxes since the giant graviton corresponds to a black D3-brane in the $\ads_5\times S^5$ background in particular with a five-form RR-flux turned on. Secondly, we are describing a brane probe that moves with constant velocity along a Killing direction. Thus, it is not a stationary solution. However, the brane probe is not accelerating in that it is moving along a geodesic. Thus, we consider what we call quasi-stationary blackfolds in the sense that they correspond to boosted stationary blackfolds. In particular we describe the conserved energy and the momentum associated with the motion and we also extend the other physical quantities as well as the variational principle to the quasi-stationary case. We remark that since a quasi-stationary blackfold is not accelerating it does not emit radiation and one can thus go beyond the probe approximation and perform a matched asymptotic expansion for the full system of the background with the brane.

For the extremal giant graviton there are two solutions for each value of $J$ in the range $0 < J < 9N/8$ and the ones with the biggest $S^3$ radius are the stable solutions that minimize the energy (here we set the radii of $\ads_5$ and $S^5$ to one). Note that this family of solutions consists of two disconnected branches, one for $0 < J \leq N$ and one for $1 < J < 9N/8$. A similar feature is found for the thermal giant graviton in that for a given temperature there are two solutions for each value of $J$ in the range $J_{\rm min} < J < J_{\rm max}$ and those with the biggest $S^3$ radius minimize the free energy and are thus thermodynamically stable. This is true both with respect to local and global stability. Thus we have shown that there are stable thermal giant gravitons in this range of $J$. Note also that the family of stable solutions again consists of two disconnected branches, one for $J_{\rm min} < J \leq N$ and one for $N < J < J_{\rm max}$.

In addition to thermalizing the ``archetypical" giant graviton, $i.e.$ a D3-brane wrapped on an $S^3$ moving inside the $S^5$ of $\ads_5\times S^5$, one can also thermalize the other possible giant gravitons. In particular, we consider in this paper also briefly the D3-brane giant graviton moving on $\ads_5$.  The thermal effects observed in this case are similar
to that of the $S^5$ case. There is again a temperature dependent lower bound on the minimum angular momentum,
with no upper bound just as for the extremal case. Moreover, the small temperature expansion of the stable branch 
of thermal giant gravitons in  this case gives the same correction to the free energy as in \eqref{Fresult}. 

This paper is built up as follows. In Sec.~\ref{sec:GGS5} we review the extremal 
giant graviton configuration for a D3-brane expanded on the $S^5$ of the $\ads_5 \times S^5$
background and set the notation for the rest of the paper. In addition to this, we also discuss a stable branch
of solutions that has not received much attention in the literature. 
 In Sec.~\ref{app:BFgaugepot} we discuss how to extend the blackfold method
 for branes in backgrounds with fluxes and derive the form of the conserved quantities
 and action  for this case. We also introduce the notion of quasi-stationary blackfolds. 
 In Sec.~\ref{sec:GGthermal} we find the thermal version of the giant
 graviton on $S^5$ using the blackfold approach and describe the solution space. 
 We then subsequently make a detailed analysis of the thermodynamic properties
 of these thermal giant gravitons, including their stability, in Sec.~\ref{sec:GGstab}. 
 In particular, we derive the low temperature correction to the free energy given
 in \eqref{Fresult}. Finally,  in  Sec.~\ref{sec:GGads} we consider very briefly the corresponding results
 for thermal giant gravitons on $\ads_5$. 
 We end with a conclusion and outlook in Sec.~\ref{sec:outlook}. 

A number of appendices is included providing further details. 
In  App.~\ref{app:stability} we present a detailed stability analysis of the extremal
giant graviton solutions, including the non-BPS branch.  The thermodynamical
blackfold action for blackfolds in background fields with non-zero fluxes and
a corresponding Smarr relation is discussed in  App.~\ref{app:thermo}.
App.~\ref{app:EOMthermal} gives a derivation of the form of the blackfold equation
of motion for thermal giant gravitons. Finally,  in App.~\ref{app:points}
we analyze the two meeting points of the thermal giant graviton solution branches.

\section{Giant graviton on $S^5$ revisited \label{sec:GGS5} } 

In this section we review the extremal giant graviton  configuration in type IIB string theory on $\ads_5 \times S^5$.
For definiteness we focus on the case in which the giant graviton sits on the $S^5$, and correspondingly also
the construction of the thermal version in Sec.~\ref{sec:GGthermal} will be confined to this case.  The case in which the
graviton is expanded on the $\ads_5$ and its thermal version is  considered briefly in Sec.~\ref{sec:GGads}. 
 
The review below will serve to set our notation and properly define the configurations that will be heated up using
the blackfold approach. At the same time we highlight that, beyond the usual 1/2 BPS solution, there is a stable branch of giant gravitons that has not  received much attention in the literature. Some unnoticed properties of this branch will
be discussed as well.

\subsection{Setup and action}

We consider ten-dimensional type IIB string theory on $\ads_5 \times S^5$ with radii $L$ and five-form flux 
\beq
\label{F5ds5}
F_{(5)} = 4 d \Omega_{(5)}/L \ , 
\eeq
 where $d \Omega_{(5)}$ is the unit volume form on the $S^5$. 
For the $S^5$ we take the parameterization
\begin{equation} \label{ds5}
\text{d}\Omega_{(5)}^2=L^2 \left [\text{d}\zeta^2+\cos^2\zeta d\phi_{1}^2 +\sin \zeta^2\text{d}\Omega_{(3)}^2 \right] \ , 
\end{equation}
where $\text{d}\Omega_{(3)}^2$ denotes the line element on $S^3$ (with coordinates $\phi_2,\phi_3,\theta$). 
The giant graviton is obtained by considering a (rotating) probe D3-brane in this space, that wraps an $S^3$ inside
the $S^5$.  
Denoting the world volume coordinates of the D3-brane as   $\{\sigma^0\equiv\tau,\sigma^i\}$, its embedding
into the background is taken to be 
\begin{equation} 
\label{D3embed}
t=\tau, \quad \phi_1=\Omega\tau, \quad \phi_2=\sigma_1, \quad \phi_3=\sigma_2, \quad \theta=\sigma_3 \spa
\zeta = {\rm const.}  \ , 
\end{equation}
while the D3-brane sits at the origin of the $\ads_5$ space.  The size of the giant graviton is thus
$r = L \sin \zeta$ and this configuration rotates with angular velocity $\Omega$ on the $S^5$, satisfying
the geometric bound $(L^2-r^2)\Omega^2\le1$. 
 The resulting induced metric on the D3-brane world volume is
 \beq
 \label{indmet}
 \gamma_{ab}d\sigma^a d\sigma^{b}=-\bk^2d\tau^2+r^2d\Omega_{(3)}^2 \ , 
 \eeq 
 where  $a=\tau,1,2,3$ runs over the world volume directions and 
 \beq
 \label{kexpression}
\bk \equiv  |k|=\sqrt{1-\Omega^2(L^2-r^2)} 
\eeq
is the norm of the rotational Killing vector satisfying $\bk \leq 1$. 

With this setup, the giant graviton is found by solving the equations of motion (EOMs) of the D3-brane
DBI action $I_{\rm DBI}$ in this background.  Defining the corresponding Lagrangian via $I_{\rm DBI} =
\int d \tau  L_{\rm DBI}$,
we have 
\begin{equation} \label{dbiL}
L_{\rm DBI} = \int_{S^3} {\cal{L}} =  -T_{\rm D3}  \int_{S^3} \left(\sqrt{-\gamma}-A_{\tau\sigma_1\sigma_2\sigma_3}\right) \ , 
\end{equation}
where $\gamma $ is the determinant of the induced metric \eqref{indmet} 
on the D3-brane and $A_{\tau\sigma_1\sigma_2\sigma_3}$ the pullback of the four-form gauge potential onto the world volume. Using the embedding above
this gives
\begin{equation}
\label{LD3}
L_{\rm DBI} =  -T_{\rm D3}\Omega_{(3)}r^3\left(\bk-r\Omega \right)  \  . 
\end{equation}  
The angular momentum and Hamiltonian are then computed as 
\begin{equation} \label{jdbi}
J=\int_{S^3}\frac{\partial\mathcal{L}}{\partial\Omega}=T_{\rm D3}\Omega_{(3)}r^3\left(\frac{\Omega(L^2-r^2)}{\bk}+r\right)
\spa H=J\Omega-\int_{S^3}\mathcal{L}=\frac{T_{\rm D3}\Omega_{(3)}r^3}{\bk} \ . 
\end{equation}
We finally note that the overall factor in all these expressions involves   $T_{\rm D3} \Omega_{(3)} = N/L^4$
where $N$ is the background flux.

\subsection{Solution branches and stability \label{sec:extsol} }

\begin{figure}[!ht]
\centerline{ \includegraphics[scale=0.3]{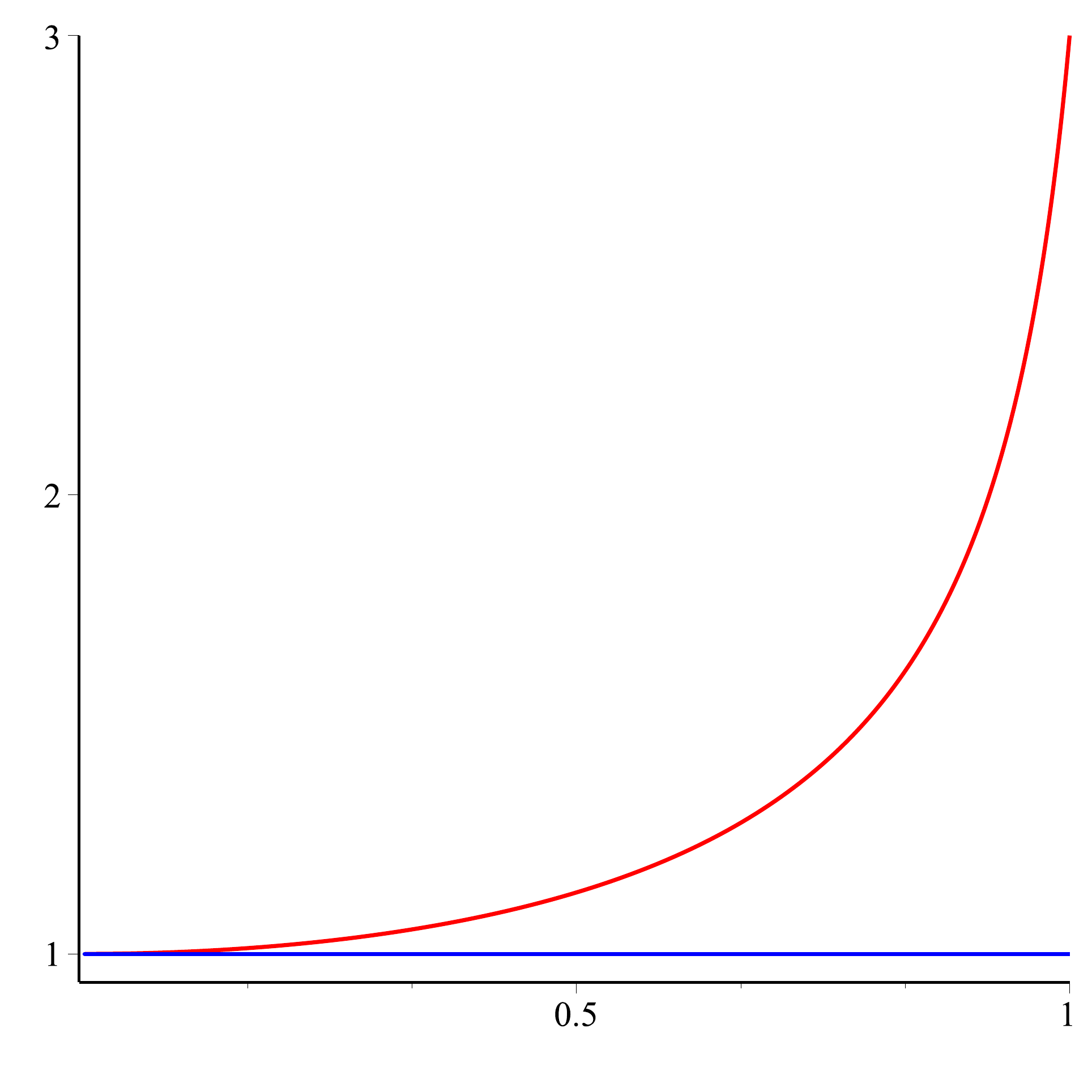} \hskip 1cm
\includegraphics[scale=0.3]{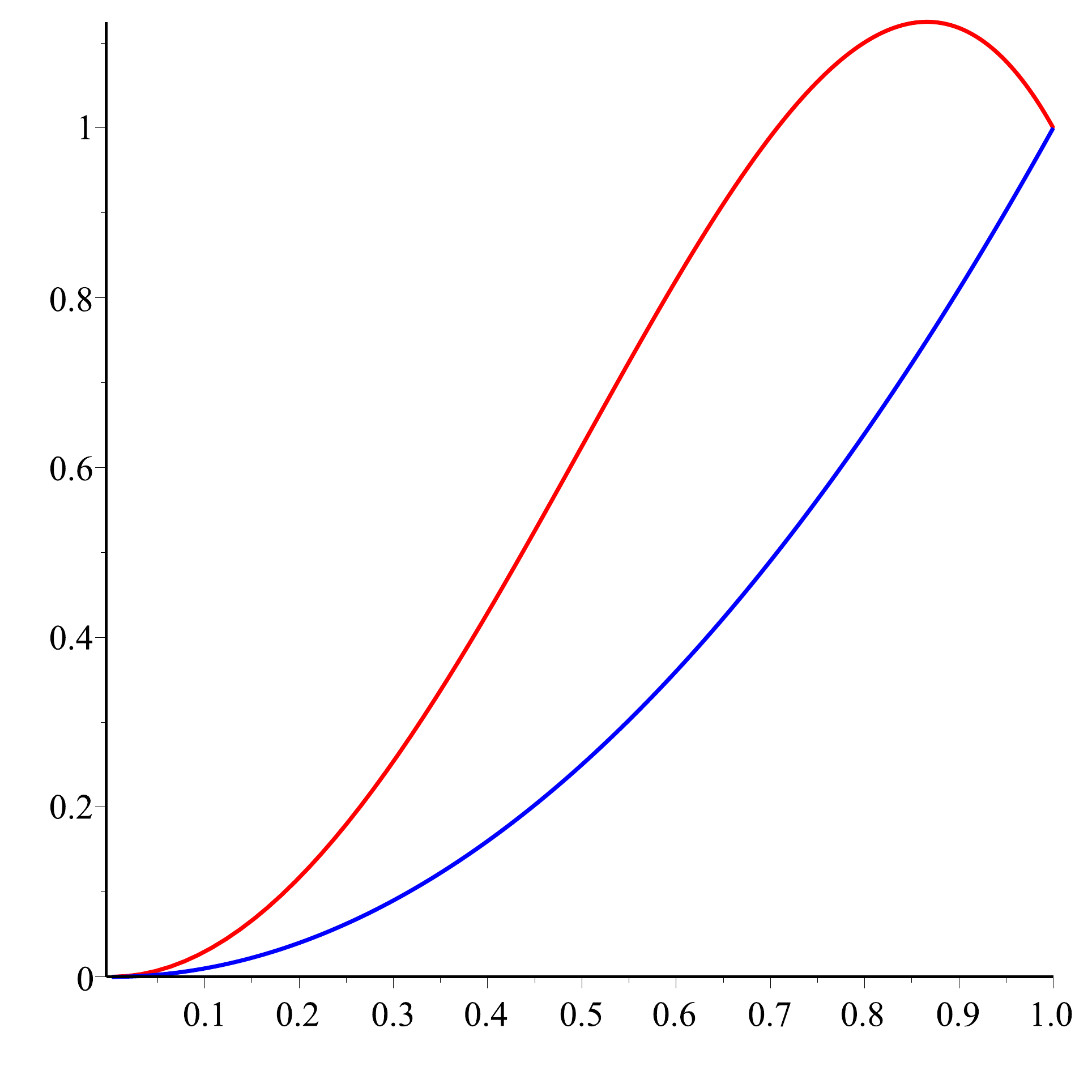} }
\begin{picture}(0,0)(0,0)
\put(210,30){ $\hat r $}
\put(45,185){ $ \hat \Omega   $}
\put(410,30){ $\hat r$}
\put(245,185){ $ \mathbf{J}  $}
\end{picture}		       
\vskip -.5cm
	\caption{$\hat \Omega \equiv \Omega L $ (left plot) and $\mathbf{J} \equiv J /N$  (right plot) versus  $\hat r \equiv r/L$ for the two solution branches of extremal giant gravitons.
	The lower ($-$) branch is blue and the upper ($+$) branch  is red. 
	} 
	\label{fig:Omegavsrext}
\end{figure}
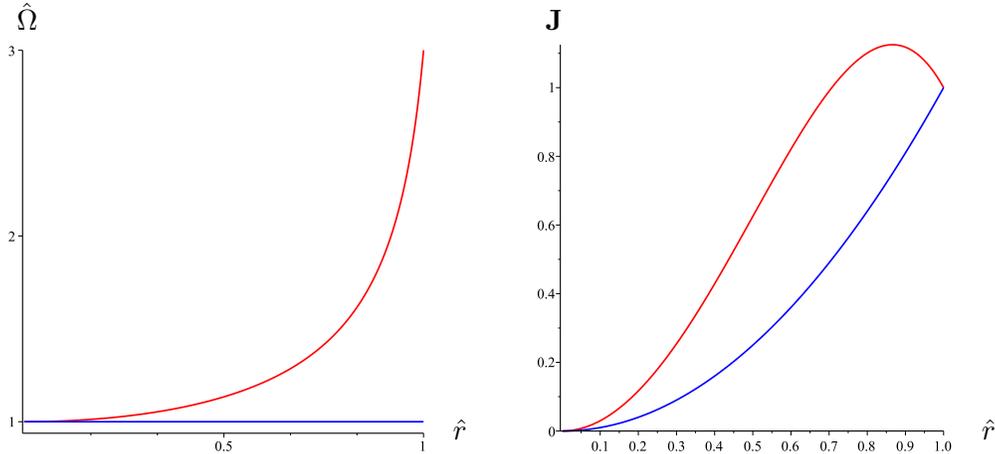

Varying the Lagrangian \eqref{LD3} with respect to $r$ we obtain the EOM
\beq \label{exl}
3-3L^2\Omega^2+4r\Omega\left(r\Omega-\sqrt{1-\Omega^2(L^2-r^2)}\right) = 0 \ . 
\eeq
This equation has two branches of solutions%
\footnote{The limit $r=0$ of these solutions describes the point-particle limit of the giant graviton, where one should be careful
in taking the limit $r\to0$ such as to obtain sensible conserved charges \cite{Grisaru:2000zn}.}
\beq
\label{Ommext} 
\bar \Omega_{-}=\frac{1}{L} 
\spa \bar \Omega_{+}=\frac{3}{\sqrt{9L^2-8r^2}} \spa (0\le r\le L )  \ , 
\eeq
which we call the lower and upper branch respectively. Note that for the upper branch we have
that  $1\leq \Omega L \leq 3$. These two solution branches are depicted in the left plot of Fig.~\ref{fig:Omegavsrext}. 
It is interesting to note that a maximal size giant graviton ($r=L$) exists in both branches with either $\bar \Omega_{-}=L^{-1}$ or $\bar \Omega_{+}=3L^{-1}$. Moreover it is also worth noticing that both branches connect to the point-particle case in the limit $r\to0$.

\vskip .7cm

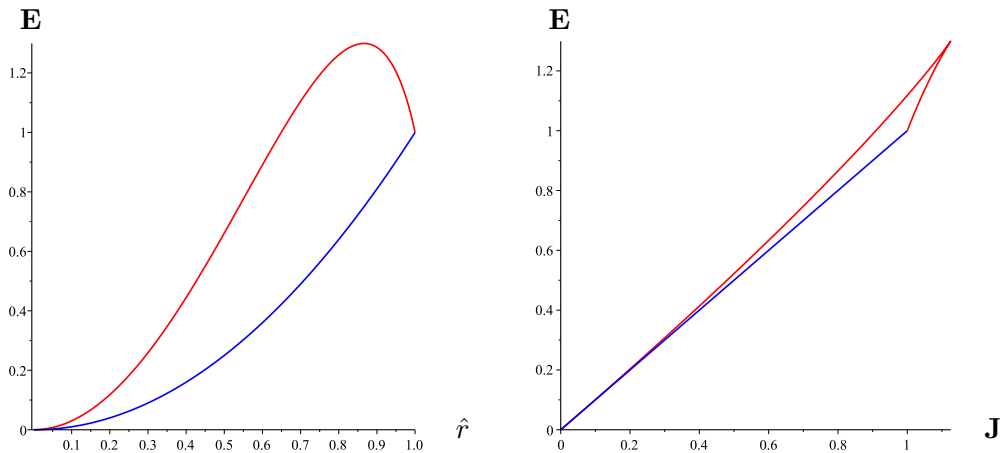
\begin{figure}[!ht]
\centerline{\includegraphics[scale=0.3]{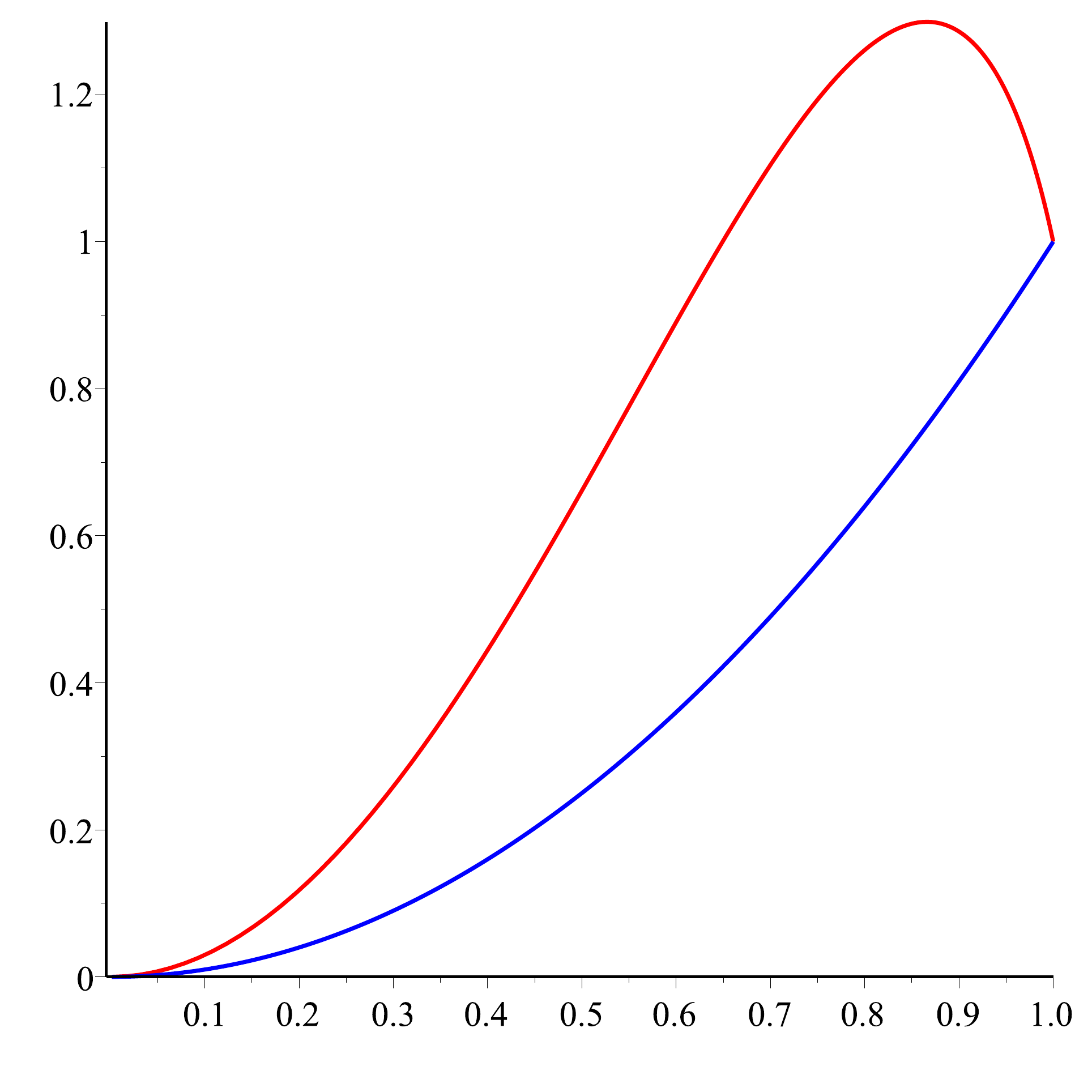} \hskip 1cm
\includegraphics[scale=0.3]{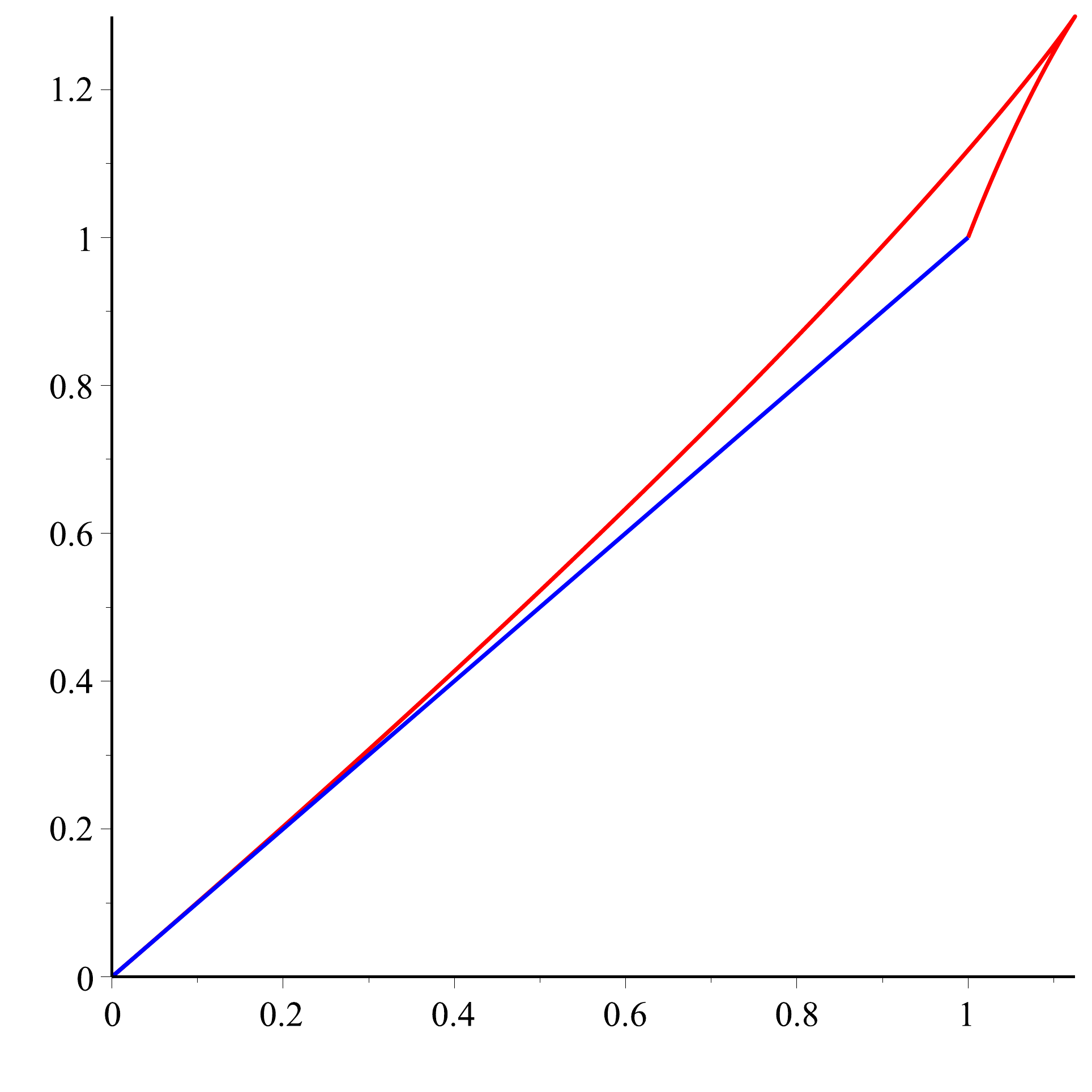} }
\begin{picture}(0,0)(0,0)
\put(210,30){ $\hat r $}
\put(45,185){ $ \mathbf{E}   $}
\put(410,30){ $\mathbf{J} $}
\put(245,185){ $ \mathbf{E}  $}
\end{picture}		
\vskip -.5cm
	\caption{$\mathbf{E} \equiv (L/N) E $  versus  $\hat r $ (left plot) and versus $\mathbf{J}$   (right plot) 
	for the two solution branches of extremal giant gravitons.} 
	\label{fig:Evsrext}
\end{figure}

To elucidate these branches and connect to a more physical parameterization  
we use \eqref{jdbi} to compute  the on-shell  angular momentum $J(\hat r)$ and energy 
$E(\hat r)$  for each of the two branches, 
where we introduce the dimensionless ratio  $\hat r = r/L$, yielding the angular momentum
\begin{equation}
\label{Jpm} 
J_{-}=  N  \hat r^2 \spa J_{+}= N \hat r^2(3-2 \hat r^2) \ , 
\end{equation}
and energy
\beq
\label{Epm}
E_{-}=  \frac{N}{L}   \hat r^2 \spa E_{+}=  \frac{N}{L} \hat r^2\sqrt{9- 8 \hat r^2 } \ , 
\eeq
on each of the two branches. 
For clarity, we have depicted these results in the right plot of Fig.~\ref{fig:Omegavsrext} 
and the left plot of  Fig.~\ref{fig:Evsrext} respectively. Here and in the following we
have also rescaled $\mathbf{J} \equiv J/N$ and  $\mathbf{E} \equiv (L/N)  E  $. 

\vskip .7cm

One observes that for each value in the range $0 \leq \mathbf{J} \leq 9/8$,
there are two possible solutions, with different values of $\hat r$.  Comparing the two corresponding values of the
energy for each of these two values of $\hat r$ (given $\mathbf{J}$), one finds that the one with highest $\hat r$
minimizes the energy. To see this more clearly, we exhibit $\mathbf{E}$ versus $\mathbf{J}$ in the right plot of Fig.~\ref{fig:Evsrext}. 
Hence we expect that the stable branch of solutions consists of the entire lower branch  (for $ 0 \leq \mathbf{J} \leq 1$
and $0 \leq \hat r \leq 1$)  together with  the part of the upper branch that has  $ 1 \leq \mathbf{J} \leq 9/8$ and $ \sqrt{3}/2 \leq \hat r \leq 1$. 
Conversely, the part of the upper  branch spanned by $ 0 \leq \mathbf{J} \leq 9/8$ and $ 0 \leq \hat r \leq \sqrt{3}/2$
will be for given $\mathbf{J}$ a local maximum of the energy.

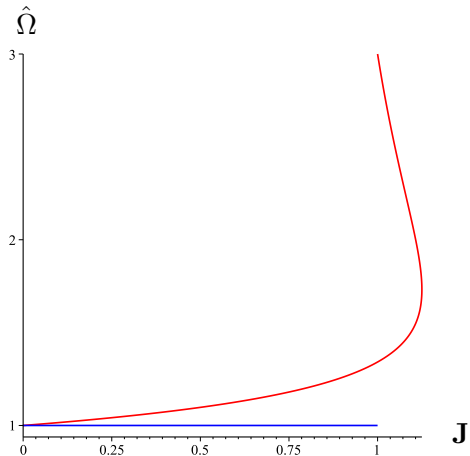
\begin{figure}[!ht]
\centerline{\includegraphics[scale=0.3]{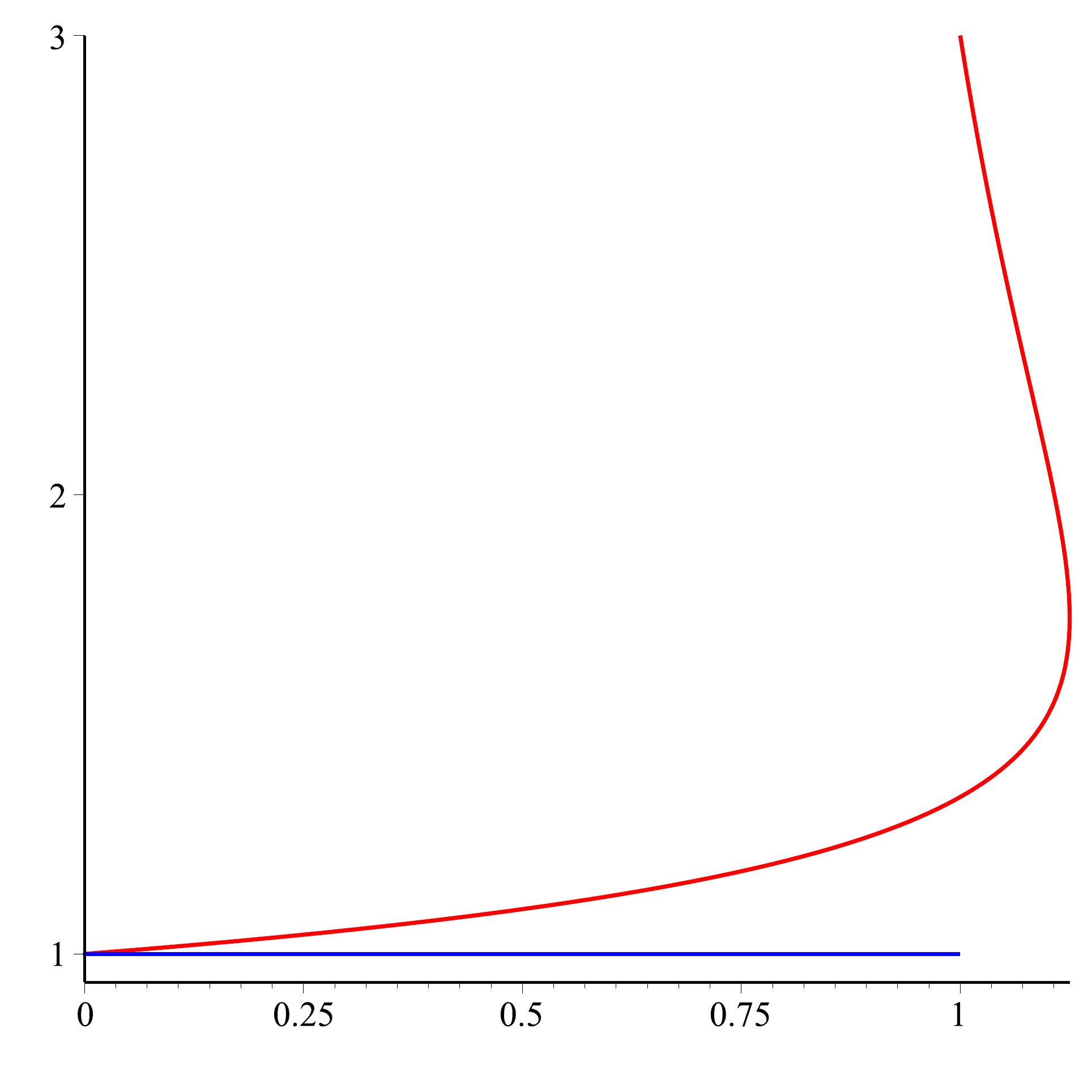}}
\begin{picture}(0,0)(0,0)
\put(310,30){ $\mathbf{J}$}
\put(145,185){ $ \hat \Omega   $}
\end{picture}		
\vskip -.5cm
	\caption{$\hat \Omega$ versus $\mathbf{J}$  for the two solution branches of extremal giant gravitons
	} 
	\label{fig:OmegavsJext}
\end{figure}

 More properly, this result on the dynamical stability can be derived by computing the off-shell Hamiltonian from  \eqref{jdbi} 
 \beq
H = \frac{ N }{L} \sqrt{ \hat r^6 +\frac{(\mathbf{J}- \hat r^4)^2 }{1-\hat r^2}} \ . 
\eeq
 Varying this with respect to $\hat r$ for constant $\mathbf{J}$ gives, as expected, the extrema $\Omega = \bar \Omega_{\pm}$ found before. 
 To see which part of the branches are stable we vary $H$ once more with respect to $\hat r$ at constant $\mathbf{J}$,
 and demand positivity, so that we are at a minimum. 
 The result is that the lower branch $\Omega= \bar \Omega_-$ is stable for all values of $\hat r$
 ($ 0  \leq \hat r \leq 1)$ and the upper branch $\Omega = \bar \Omega_+ $ is stable for  $ \sqrt{3}/2 \leq \hat r \leq 1 $.
 This is in accord with the arguments of the previous paragraph (see also App.~\ref{app:stability} where
 the same conclusion is obtained from  a more detailed stability analysis that  includes time derivatives of the radial coordinate). Finally, we note that the point $\hat r = \sqrt{3}/2$ where the upper branch becomes unstable
 can also be seen as a turning point in a plot of $\Omega$ as a function of $J$ (see Fig.~\ref{fig:OmegavsJext}).

 The main motivation of our review above and the various plots that are presented is that they will be instructive
 to illustrate the new features that appear when constructing and analyzing the thermal giant gravitons in Secs.~\ref{sec:GGthermal}  and \ref{sec:GGstab}.

\vskip .7cm

 Having established which solutions are stable, we now turn to their
 physical relevance.  First we note that they are distinguished by the angular momentum $\mathbf{J}$. 
 In terms of $\hat r$ they coexist when $ \sqrt{3}/2 \leq \hat r \leq 1$, but from \eqref{Epm} we easily see that
 the energy on the lower
 branch is lower than that of the upper branch, for given $\hat r$ in that range, except when $\hat r=1$ where they
 have the same energy. 
  The lower branch is the usual 1/2 BPS branch extensively considered in the literature, 
  and from \eqref{Jpm}, \eqref{Epm} we immediately see the BPS condition $E_- =  J_-/L$. 
 The other stable solution which is part of the upper branch was noted in \cite{Grisaru:2000zn} (see in particular Figs.~1 and 2 of that reference), 
 but has otherwise largely been ignored. First of all this branch is not connected to the point particle
 case as a stable configuration since local stability requires  $ \sqrt{3}/2 \leq \hat r \leq 1$. 
  Furthermore, while it is a local minimum of the energy it is 
 not a global one, so it is a metastable configuration and has $E_+ \geq J_+/L$ where the bound
 is saturated for $\hat r =1$.  This thus raises the question whether  this configuration indeed preserves 1/2 of the supersymmetries.
 By repeating the steps of Sec.~3 of \cite{Grisaru:2000zn}, we have verified that this is indeed the case.  The main point here is that
 the $\Omega$-dependent terms in this computation vanish at $\hat r=1$. 
   So we see that at $\hat r =1$ we can have either $\bar \Omega_- = 1/L$ or $\bar \Omega_+ = 3/L$,
 both satisfying the same BPS bound and both being supersymmetric. In particular, we cannot distinguish these configurations
 according to their energy.%
 \footnote{They also satisfy the zero temperature limit of a general Smarr relation that is derived
 in App.~\ref{app:thermo}.}
 In the first case the center of mass is rotating at the speed of light while in the second the center of mass is rotating at a superluminal velocity. However this should not be an argument for discarding the latter solution since the center of mass being a geometrical construction can be moving with superluminal velocities as long as every point on the brane is subluminal. 
 The existence of these two BPS configurations at $\hat r=1$, arising from two distinct solution branches 
 raises the question of what the dual CFT interpretation is of the one connected to the non-BPS branch, and we 
 briefly comment on this in the next subsection. 
 
 The primary purpose of this paper though  is to use the blackfold approach to find the thermal versions of the 
 extremal configurations reviewed in this section. Concretely, we will switch on a temperature and examine what
 happens with the {\it entire} solution branch depicted in Figs.~\ref{fig:Omegavsrext}, \ref{fig:Evsrext} and \ref{fig:OmegavsJext} irrespective of their stability properties and subsequently identify the stable branches.  As it will be discussed in more detail below, in the blackfold approach
 one obtains solutions in the supergravity (closed string) regime. In this sense the entire solution branch
 (stable or not) is a valid approximate analytical solution of the type IIB supergravity EOMs, at least in the 
 pertubative regime where the approach is valid.  Moreover, just as a subset of the branches in the extremal case is locally stable, we will likewise see  that a corresponding subset is dynamically stable when the system
 is considered at non-zero temperature. 

\subsection{CFT dual and correlation functions} 

The CFT dual operator of a single point-like graviton is a chiral primary of the form
\beq \label{po}
\mathcal{O}=\text{Tr} \, Z^{J} \  , 
\eeq
with $J$ the angular momentum on the $S^5$ and $Z$ a complex scalar field. 
Standard computations have shown that their two- and three point functions match exactly on both gauge and string theory sides provided $J$ is small. 

If $J \gg N/\sqrt{\lambda}$ the correct description is in terms of a giant graviton. The dual gauge theory operator $\CO_{\rm gg}$ of the giant graviton is no longer given by \eqref{po} and arguments based on symmetry 
(which only really apply close to $r=L$)  imply that it must be replaced by a Schur polynomial operator of the form \cite{Corley:2001zk} (see also \cite{Balasubramanian:2001nh})
\beq\label{so}
\CO_{\rm gg} \sim
\chi_{R}(Z)=\frac{1}{J!}\sum_{\sigma\in S_{J}}\chi_{R_J}(\sigma)Z^{i\sigma(1)}_{i_1}...Z^{i\sigma(J)}_{i_J} \ , 
\eeq
where $Z$ is a complex matrix, $R_n$ denotes an irreducible representation of $U(N)$ described in terms of a Young tableau with $J$ boxes.

As explained above, there is another (upper) branch of giant gravitons  which is $1/2$ BPS at $r=L$ 
in the large $J$ limit with the same quantum numbers as the lower branch. 
We speculate that there exists another 1/2 BPS Schur polynomial operator in the CFT at $J=N$ that is
distinct from the Schur polynomial relevant to the usual (lower) BPS branch and which is dual to the upper branch of giant gravitons at $r=L$. We present indications of this below.

\subsubsection*{Two-point correlation functions}
As an explicit check of the statement above, we now compute the two-point function for the CFT operator dual to the 
$r=L$ point on the upper branch, showing that it has the same properties  as the $r=L$ solution of the lower branch.
It is easiest to do the computation simultaneously for both branches.
Our method is based on the general prescription, reviewed in \cite{Janik:2010gc}, for computing two-point correlation functions for massive  (or light) particles moving in a background spacetime. 

The giant graviton is  a brane, not a particle, however as seen from the $\ads_5$ part it is a point-particle with a certain mass \cite{Caldarelli:2004yk}. This can be seen by introducing motion in the $\ads_5$ part, i.e. introducing the dependence
$x^\mu (\tau)$,  $\mu = 0 \ldots 4$ on the  coordinates of $\ads_5$ with metric $G_{\mu \nu}$.
Following \cite{Caldarelli:2004yk} one can then show that the DBI action can equivalently be written
as
\beq
I_{\rm DBI}=\frac{1}{2}\int d\tau \left(\frac{G_{\mu\nu} \dot{x}^{\mu} \dot{x}^{\nu}}{e}+\frac{\Omega^2(L^2-r^2)}{e}-m^2e+
m^2r\Omega\right) \ , 
\eeq
where we have defined $m= N r^3/L^4$ and $e$ is an einbein which acts as a Lagrangian multiplier.  
Using \eqref{jdbi} we can eliminate $\Omega$ in favor of $J$ and arrive at the action
\beq
I = \frac{1}{2}\int d\tau \left(\frac{G_{\mu\nu} \dot{x}^{\mu} \dot{x}^{\nu}}{e}+e M^2 \right) \ , 
\eeq
where we have defined
\beq
\label{massM}
M=\sqrt{\frac{J^2-L^2m^2}{L^2-r^2}} \ . 
\eeq
However, to arrive at the interpretation that from the $\ads_{5}$ perspective the giant graviton is a massive point particle
moving along a timeline geodesic, one should take into account that $J$ must be conserved along any path.
Hence, one should consider the Routhian ${\cal{R}}$ which is obtained by doing a Legendre transformation in the cyclic coordinates. In this case it coincides with the Hamiltonian, and hence we find
\beq
\label{Routhian} 
{\cal{R}} = H =\Omega J - L =-\frac{1}{2}\left(\frac{G_{\mu\nu} \dot{x}^{\mu} \dot{x}^{\nu}}{e}-e E^2\right) \ , 
\eeq
where $E$ is the  on-shell energy \eqref{Epm} for each of the branches.
So we find that from the AdS$_{5}$ perspective the giant graviton is a point particle with mass $E$.
Following \cite{Janik:2010gc}, wee can now compute the two-point function using the Routhian 
\beq
G(0,\epsilon;x,\epsilon)=e^{-{\cal{R}} } \sim \left(\frac{|x|}{\epsilon}\right)^{-2E_{\pm} } \ , 
\eeq
showing for both branches equality of the anomalous dimension and the energy.
We thus conclude that the anomalous dimension of the operator is equal to the energy for both branches,
thus giving strong indication of being in both cases a Schur polynomial at the $r=L$ point.

It is important to note that the correct result is reproduced here using the Routhian, and not the action,
as was also advocated recently in \cite{Bak:2011yy}.  Indeed, evaluating the quantity $M$ in  \eqref{massM} for each of the
solution branches found in subsection \ref{sec:extsol} one finds%
\footnote{In Ref.~\cite{Bissi:2011dc} the action was used to compute the two-point function, but since this computation was
for the lower (1/2-BPS) branch, for which the terms conspire to give $M_- = E_-$, this still gives the correct result.}
  \beq
M_-= \frac{N}{L} \hat r^2  =E_{-} \spa M_+=  \frac{N}{L} \hat r^2 \sqrt{9-4 \hat r^2}  \neq E_+ \ , 
\eeq
as compared to the energies given in \eqref{Epm}.

\subsubsection*{Three-point correlation functions}

To  gain further insight into the nature of the new state at $r=L$ one may consider 
the three-point correlation function between one point particle and two giant gravitons. For the lower branch this
analysis was performed in \cite{Bissi:2011dc}.  The procedure consists in analyzing the supergravity modes  which describe fluctuations in the Euclidean D-brane action of the metric and 4-form potential,  which are dual to chiral primary operators with R-charges  in the  $\mathcal{N}=4 ~\text{SYM}$ theory. 
The resulting three-point function structure constant for the maximal size 1/2-BPS giant graviton was found to be zero
in agreement with the gauge theory side. Following the same steps for the upper branch $r=L$ state gives zero as well,
since one can check that in that case the result is independent of $\Omega$. This provides further confirmation
that the gauge theory description of the upper branch $r=L$ state is  a Schur polynomial. It would be very interesting
to calculate this three-point function more generally for the entire (non-BPS) upper branch, but this is beyond the
scope of the present paper. A naive application of the ideas mentioned above does not give sensible results, so
perhaps one should use the Routhian rather than the action and/or introduce an appropriate cutoff to regularize the
divergent integrals.

\section{Blackfold method for branes in flux backgrounds \label{app:BFgaugepot}  }

In this section we briefly discuss how to extend the blackfold method for branes embedded in a background with fluxes. 
Moreover, we present a slight generalization of the previously obtained results for stationary blackfolds to blackfolds that
are ``boosted stationary" blackfolds, with the boost  along a Killing direction of the background.  
These ingredients are relevant to the application of the blackfold method in Secs.~\ref{sec:GGthermal}-\ref{sec:GGads} to construct and analyze  localized thermal giant gravitons based on black D3-branes wrapping an $S^3$ in the
$\ads_5 \times S^5$ background. 

Our conventions follow \cite{Emparan:2009at,Emparan:2011hg} (see also \cite{ Armas:2011uf}).  In particular, space-time coordinates are denoted by $x^\mu$, brane world volume coordinates by  $\sigma^a$. Furthermore, the brane embedding in the background space-time with metric $g_{\mu \nu}$ is denoted
by  $X^\mu(\sigma)$, so that $\gamma_{ab} = g_{\mu \nu}\partial_a X^\mu \partial_b X^\nu $ is the induced metric on the brane world volume.
We will also need the orthogonal projector $ {\perp^\mu}_\nu$ defined by $g^{\mu\nu} = h^{\mu\nu} + \perp^{\mu\nu}$ where the tangential projector is $h^{\mu\nu}=\gamma^{ab} \partial_a X^\mu \partial_b X^\nu$.

\subsection{Blackfold EOMs \label{sec:BFeom} }

Consider a $p$-brane embedded in a $D$-dimensional background. In the (probe) blackfold approximation we regard this brane as infinitely thin and thus having a localized stress-tensor
\begin{equation}
\label{That}
\hat{T}^{\mu\nu}(x)=\int d^{p+1} \sigma \sqrt{-\gamma} \frac{\delta^{(D)}(x-X(\sigma))}{\sqrt{-g}}   T^{\mu\nu}(\sigma) \ , 
\end{equation}
where the integral is over the world volume ${\cal{W}}_{p+1}$ of the brane. 
In the following we furthermore consider the background to have a $(p+2)$-form flux  $F_{(p+2)} = d A_{(p+1)}$ under which the $p$-brane is electrically charged with charge $Q_p$. The charged $p$-brane therefore has the $(p+1)$-form current
\begin{equation}
\label{Jhat}
\hat{J}_{(p+1)}(x)=\int d^{p+1}\sigma \sqrt{-\gamma} \frac{\delta^{(D)}(x-X(\sigma))}{\sqrt{-g}} J_{(p+1)}(\sigma) \ , 
\end{equation}
with
\begin{equation}
J_{(p+1)} = Q_p \omega_{(p+1)} \spa \omega_{(p+1)} = \sqrt{-\gamma} \, d\sigma^0 \wedge \cdots \wedge d\sigma^p \ . 
\end{equation}
The EOMs for the charged brane probe in a background flux are
\begin{equation}
\label{hatEOM}
\nabla_\mu \hat{T}^{\mu\nu} = \frac{1}{(p+1)!} F^{\nu \rho_1 \cdots \rho_{p+1}} \hat{J}_{\rho_1 \cdots \rho_{p+1}} \ , 
\end{equation}
\begin{equation}
\label{hatcurrentcon}
\nabla_\mu \hat{J}^{\mu \nu_1 \cdots \nu_p} = 0 \ , 
\end{equation}
where the righthand side of \eqref{hatEOM} includes  the generalized Lorentz force. 
By projecting these equations on directions parallel and perpendicular to the world volume of the brane,
we get a set of intrinsic and extrinsic equations respectively. 
The intrinsic equations include $D_a J^{a a_1 \ldots a_p} = 0$, which correspond to charge
conservation as well as\footnote{In the case of a submanifold with boundaries (with unit normal vector $\hat{n}_{a}$) these must be supplemented by the boundary conditions
$ T^{ab}\hat{n}_{b}|_{\partial \mathcal{W}_{p+1}}=0$ and $J^{a}\hat{n}_a|_{\partial \mathcal{W}_{p+1}}=0$. } 
\beq
\label{inteq}
  D_a T^{ab} =0 \ , 
\eeq
under the assumption that the background gauge-potential is constant along
the world-volume of the brane, so that there are no forces induced on the world volume. This will be the case in our applications. 
The extrinsic equation for the brane embedding is (see e.g. \cite{Carter:2000wv,Emparan:2009at,Emparan:2011hg})
\begin{equation}
\label{gen_extr_eq}
T^{ab} K_{ab}{}^\mu = \frac{1}{(p+1)!}  {\perp^\mu}_\nu F^{\nu \rho_1 \cdots \rho_{p+1}} J_{\rho_1 \cdots \rho_{p+1}} \ , 
\end{equation}
where
\begin{equation}
\label{Kdef}
K_{ab}^{\ \ \mu}=\perp^{\mu}_{\ \lambda}\left\{\partial_a\partial_b X^\lambda+\Gamma^{\lambda}_{\nu\rho}
\partial_a X^\nu \partial_b X^\rho \right\}
\end{equation}
is the second fundamental tensor. 
 
We note here that the EOMs for extremal F-strings, D/M-branes in backgrounds with appropriate non-zero
flux and constant dilation field are of the form  \eqref{gen_extr_eq} with appropriate $T_{ab}$.  
In particular, the D-brane equation of motion coming from the DBI action (including the WZ term), is of
the form \eqref{gen_extr_eq} with  $T_{ab} = -T_{Dp} \gamma_{ab}  $ in the case of zero world volume gauge field. 
In the blackfold approach this  energy momentum tensor
is replaced by the effective stress tensor of the corresponding non-extremal brane solution of supergravity. 
In the leading order approximation this is that of a perfect fluid in local thermodynamic equilibrium,
 with  equation of state determined by the brane solution.  Solution of the intrinsic equation \eqref{inteq} requires 
 in that case that the fluid velocity $u^a = k^a /|k | $  is aligned with a timelike world volume Killing field 
  and following \cite{Emparan:2009at}
 we assume that  this  Killing field $k_a = \partial_a X^\mu (\sigma) k_\mu $ can be pushed forward to a corresponding timelike background Killing field $k^\mu$.     This will be the context in which we will derive conserved quantities below.

\subsubsection*{Stationary and quasi-stationary solutions}

From the above, it follows that in general the normal to the spacelike hypersurface $\CB_p$ of the blackfold
world volume $\CW_{p+1}$ need  not be parallel to the
generator of asymptotic time translations. i.e.  one can have
\beq  
\label{deltau}
\partial_\tau = a  \xi + b \chi \spa 
\eeq
where $\xi$ corresponds to the canonically normalized generator of time translation 
and $\chi$ that of a spatial $U(1)$ isometry of the background. 
 However,  with regard to stationarity there is an important further distinction 
to be made, depending on whether:%
\footnote{For simplicity we restrict to these two possibilities, but we note that when $\chi$ is not a world volume Killing vector,
it could also have some components along the world volume. E.g.   $\chi$ could be a linear combination of  a woldvolume Killing vector and a perpendicular component along an isometry of the background.}
\begin{itemize}
\item[i)] $\chi$ is also a worldvolume Killing vector 
\item[ii)] $\chi$ is perpendicular to the world volume (and hence not a world volume Killing vector) 
\end{itemize}

In case i)  the blackfold world volume  does not break the isometries $\xi, \chi$ of the background and in particular
the conserved quantities associated to these are also conserved for the entire solution consisting of 
background with the blackfold in it.%
\footnote{This is for example the case for the configurations considered in  Ref.~\cite{Camps:2008hb}  using the blackfold
method.}  The resulting solutions are stationary blackfolds.  
 In case ii), which is the one relevant for the present paper, the blackfold  world volume only
preserves a particular combination of the isometries $\xi, \chi$. As a result the conserved quantities
 associated to these are of a different nature,  namely they  refer to quantities for the probe blackfold 
 in the background space time but are not seperately conserved for the background including the blackfold. 
  Only an appropriate linear combination is conserved, according to 
\eqref{deltau}.  In particular,  the conserved quantity generated by $\xi$ should be thought of as the total energy 
$E$ (so not the rest mass of the object) and the quantity generated by $\chi$  as the  transverse momentum $J$ corresponding to
the boost.  So we see that in this case the blackfold is  transversely boosted along a Killing isometry of the background, 
and hence it should be viewed as a "boosted stationary" solution. We will refer to this below as quasi-stationary, since we still have that, seen from the world volume,  the blackfold is independent of time.
 This is in fact precisely what happens  for the localized giant
graviton, since in that case $\chi$ lies in a direction perpendicular to the world volume. 
It is important to note that since the quasi-stationary blackfold is not accelerating it does not emit radiation and one can thus go beyond the probe approximation and perform a matched asymptotic expansion for the full system of the background with the brane.

\subsection{Conserved charges \label{sec:charges}}

We now write down the expressions for the conserved charges
corresponding to the asymptotic generators $\xi$ and $\chi$ for these quasi-stationary blackfolds in flux backgrounds.
Note that the results below can also be used for stationary blackfolds in flux backgrounds. 

For any Killing vector field (KVF) $k$ of the background we have by definition that the Lie derivative along $k$ of the $(p+2)$-form $F=F_{(p+2)}$ 
is zero $\CL_k F = 0$. Since $dF = 0$ we find that $0 = dF = i_k dF + d (i_k F) = d (i_k F)$ where $i_k$ means the contraction with the KVF $k$. Picking a gauge in which $\CL_k A =0$ we see indeed that $0 = \CL_k A = i_k F  + d ( i_k A )$ thus $i_k F = - d (i_k A )$. Thus, in this gauge the $(p+1)$-form $i_k F$ has the $p$-form potential $i_k A$. Using this with the EOM \eqref{hatEOM} and current conservation \eqref{hatcurrentcon} we see that
\begin{eqnarray}
p ! \ \nabla_\mu ( \hat{T}^{\mu \nu} k_\nu ) &=& \frac{1}{p+1} k^\nu F_{\nu \rho_1 \cdots \rho_{p+1}} \hat{J}^{\rho_1 \cdots \rho_{p+1}} =
- \nabla_{[ \rho_1} (i_k A)_{\rho_2 \cdots \rho_{p+1}]} \hat{J}^{\rho_1 \cdots \rho_{p+1}} \nn \\ 
&=& - \nabla_{\rho_1} (i_k A)_{\rho_2 \cdots \rho_{p+1}} \hat{J}^{\rho_1 \cdots \rho_{p+1}} = - \nabla_{\mu} (  A_{\nu \rho_1 \cdots \rho_{p}} \hat{J}^{\mu \rho_1 \cdots \rho_{p}} k^\nu   )  \  . 
\end{eqnarray}
Thus we obtain the conserved current 
\beq
\label{conscur}
j^{\mu}_k= \Big( \hat{T}^{\mu\nu}_{p}    +  \frac{1}{p!} 
A^{\nu}{}_{ \rho_1 \cdots \rho_{p}} \hat{J}^{\mu \rho_1 \cdots \rho_{p}}   
\Big) k_{\nu}    \ , 
\eeq
and a conserved charge
\beq
\mathcal{Q}_{k}=\int_{\Sigma}dx^{D-1}\sqrt{g_{\rm space}}j^{\mu}_{k}n_{\mu} \ , 
\eeq
where $g_{\rm space}$ only involves the spatial components of the background metric, 
defined by the slice $\Sigma$ of constant $x^0 = t$ and $n_\mu$ is the unit normal of $\Sigma$. 
 Inserting the conserved current \eqref{conscur}  and using the form of the stress tensor \eqref{That} and current  \eqref{Jhat}
we can do the $\delta$-function integrals to reduce to integrals over $\CB_p$.  
On the world-volume we choose the static gauge in which $X^0 = \tau  = \sigma_0$. 
 Restricting  to {\it static} backgrounds, which is sufficient for the applications in this paper,  
 we have  $\sqrt{-g}=\sqrt{-g_{00}}\sqrt{g_{\rm space}}$ and since we assume that the pullback 
$\xi^{a}$ is hypersurface orthogonal, we split $\sqrt{-\gamma}=\sqrt{-\gamma_{\tau \tau}} d V_{p}$. Then, integrating out the delta function we obtain for the conserved charge
\beq
\label{Qk} 
\mathcal{Q}_{k}=\int_{\CB_{p}}dV_{(p)}\gamma_\perp^{-1} \left[ T^{\mu\nu}  + \mathcal{V}^{\mu\nu}  
\right ] n_\mu k_\nu \vert_{x^{\mu}=X^{\mu}} \ . 
\eeq
Here we have defined
\begin{equation}
\label{Vdef}
\mathcal{V}^{\mu\nu} 
\equiv \frac{1}{p!} \thinspace A^{\nu}_{\ \thinspace \mu_1\cdots\mu_{p-1}}J^{\mu \mu_1\cdots\mu_{p-1}} \ , 
\end{equation}
and the transverse Lorentz contraction factor  $\gamma_\perp$ is given by 
\beq
\label{gamma}
\gamma_\perp   \equiv  \frac{R_0}{\rho_0} \ , 
\eeq
where $R_0 \equiv \sqrt{-g_{tt}}$ and $\rho_0 \equiv \sqrt{-\gamma_{\tau \tau}}$.  

Now, we use the result \eqref{Qk} to write down conserved charges 
 corresponding to the background Killing vectors $\xi $ and $\chi$ 
\beq 
\label{Mres2}
E =\int_{\mathcal{B}_{p}} dV_{(p)}
\gamma_\perp^{-1} \left[ T^{\mu\nu}  + \mathcal{V}^{\mu\nu}  
\right ] n_\mu \xi_\nu 
\spa 
J =-\int_{\mathcal{B}_{p}} dV_{(p)}
\gamma_\perp^{-1} \left[ T^{\mu\nu}  + \mathcal{V}^{\mu\nu}  
\right ] n_\mu \chi_\nu  \ , 
\eeq
which are the energy $E$ and momentum $J$ of the quasi-stationary  blackfold moving with constant velocity in the background along an isometric direction. 
The interpretation of these conserved charges in this case is most easily seen by analogy
with a probe particle moving in a time-independent background along a Killing direction. 
In that case, it follows from standard analytical mechanics that the energy $E$ and momentum $P$ of the probe particle are conserved as long as the object moves with constant velocity. Likewise for the blackfold, as long as we are working 
in the leading order probe approximation where the laws of physics do not involve the internal degrees of freedom, the conservation of $E$ and $J$ relies entirely on properties of the background and not those of the brane. 
In particular, these quantities are conserved for the quasi-stationary blackfold probe just as they are for the DBI
D-brane probe.  The expressions \eqref{Mres2} will thus provide us the conserved quantities
for the thermal giant graviton constructed in Sec.~\ref{sec:GGthermal}, where we will also see that they reduce
to the correct DBI quantities in the extremal limit.

\subsection{Action and thermodynamics}

Finally, we discuss here the action that describes the quasi-stationary blackfolds introduced above. 
Again, we emphasize that the considerations of this section can also be applied to the stationary case. 

When the intrinsic EOMs are solved as explained in Sec.~\ref{sec:BFeom}  the remaining 
extrinsic EOMs \eqref{gen_extr_eq}  can be shown to follow from the Lorentzian action
\begin{equation}
\label{BFact}
I =  \int_{\CW_{p+1}} \big\{ \omega_{(p+1)} P + Q_p {\cal{P}} [A_{(p+1)} ] \big\} \ , 
\end{equation}
where $P$ is the local pressure of the blackfold and $  {\cal{P}} [A_{(p+1)}]$  the pull-back of the background
gauge potential to the worldvolume. This natural generalization of the blackfold action to include
background fluxes, easily follows form the derivations presented in Refs.~\cite{Emparan:2009at,Emparan:2011hg}.
For a given set of global parameters $(T,\Omega,Q_p)$ the (quasi)-stationary solution to the blackfold equations
of motion is exactly the one that extremizes the action $\delta I =0$.   Note also that for
extremal D-branes this action reduces to the DBI action plus WZ term.

Following the arguments in Refs.~\cite{Emparan:2009at,Emparan:2011hg}, we show in App.~\ref{app:thermo} that
the Euclidean action $I_E$ obtained by Wick rotating \eqref{BFact}, is again equivalent to the thermodynamic action%
\begin{equation}
\label{actE}
\frac{I_E }{\beta} =  F = E-\Omega J-TS \ . 
\end{equation}
Here the total entropy $S$ of 
 the blackfold is obtained in the usual manner by integrating the temporal part of the entropy current
  $su^\mu$ over $\mathcal B_{p}$, so that  
\begin{equation}
\label{BFentropy} 
S=\int_{\mathcal B_{p}}\gamma_\perp^{-1} su^\mu n_\mu \ , 
\end{equation}
where the integrand is multiplied by the appropriate Lorentz factor $\gamma_\perp$ defined in \eqref{gamma}.
Consequently we find that (at fixed $Q_p$), the extrema of the action obey the first law of thermodynamics 
$ \text{d}E=\Omega\text{d}J+T\text{d}S$. We finally note that a corresponding Smarr relation for the
thermodynamic quantities is derived in 
App.~\ref{app:thermo} as well.


\section{Construction of finite temperature giant graviton on $S^5$ \label{sec:GGthermal} } 

In this section we find a thermal version of the giant graviton consisting of a D3-brane wrapped on a 3-sphere moving on the 5-sphere of $\ads_5\times S^5$ (reviewed for the extremal case in Sec.~\ref{sec:GGS5}). This is done using the blackfold approach in the probe approximation (as reviewed in Sec.~\ref{sec:BFeom}). 
 
\subsection{Equation of motion for thermal D3-brane giant graviton}

In the leading order blackfold approximation to black branes we use the general extrinsic equation \eqref{gen_extr_eq} with the stress-tensor $T_{ab}$ being that of a black brane. In particular, in the present case we want to consider  black D3-branes in the $\ads_5\times S^5$ background. Specifically, the stress-tensor of black D3-branes corresponds to that of a four-dimensional fluid tensor of the form $T_{ab} = (\epsilon + P) u_a u_b + P \gamma_{ab}$ with $u_a$ being the four-velocity and energy and pressure given as
\begin{equation}\label{thermoquantities}
\epsilon = \mathcal T s - P  \spa P = -\mathcal{G}\left(1+4\sinh^2 \alpha\right) \spa \mathcal T s=4 \mathcal G \spa  \mathcal G \equiv  \frac{\pi^2}{2} T_{\rm D3}^2  r_0^4
\end{equation}
where the local temperature $\mathcal T$ and entropy density $s$ for the black D3-brane are
\begin{equation}
\label{TandS}
\mathcal T = \frac{1}{\pi r_0 \cosh \alpha} \spa s = 2\pi^3 T_{\rm D3}^2 r_0^5 \cosh \alpha
\end{equation}
The parameters of the black D3-brane stress tensor and thermodynamics are thus $r_0$ and $\alpha$. 
The black D3-brane furthermore has the 4-form charge current 
\begin{equation}
\label{D3current} 
J_{(4)} = Q d\tau d\sigma^1 d\sigma^2 d\sigma^3  \spa {Q}=4\mathcal G \sinh \alpha \cosh \alpha = N_{\rm D3} T_{\rm D3}
\end{equation}
where $Q$ is the charge density and
$N_{\rm D3}$ is the number of coincident D3-branes and $T_{\rm D3}= 1/((2\pi)^3 g_s l_s^4)$ is the D3-brane tension. Note that we have the relation $\Omega_{(3)} T_{\rm D3} = N / L^4$ where $N$ and $L$ are the magnitude of the flux and the radius in $\ads_5 \times S^5$, respectively. 

The giant graviton that we wish to thermalize is that of a D3-brane wrapped on a 3-sphere moving on the 5-sphere of $\ads_5\times S^5$. This giant graviton solution, as studied using the DBI action for the extremal D3-brane, was reviewed above in Sec.~\ref{sec:GGS5}. In particular, the metric on the  5-sphere  is given by \eqref{ds5} and the 5-form flux in  \eqref{F5ds5}.
 We now use the properties \eqref{thermoquantities}-\eqref{D3current} of black D3-branes
 to study the case where the D3-brane is put at finite temperature, hence we are studying what can be called a thermal giant graviton in the form of a black D3-brane wrapped on a 3-sphere moving on the 5-sphere of $\ads_5\times S^5$. We take the same ansatz for the embedding \eqref{D3embed} as used for the extremal D3-brane. 
 Using all this information (see App.~\ref{app:EOMthermal} for details of the derivation)%
\footnote{Note that the worldvolume velocity field is given by $u_\tau = 1/\bk$ so on the worldvolume the D3-brane is static,
but the push-forward of this vector to the background gives the vector field $u^\mu \partial_\mu \sim \partial_t  + \Omega \partial_\phi$, so this is a quasi-stationary blackfold as explained in Sec.~\ref{sec:BFeom}.}
 with Eq.~\eqref{gen_extr_eq} we arrive after some algebra at the following EOM
\begin{equation}
\label{exteq}
\Omega^2r^2\left(1-\CR_1(\alpha) \right) +3\bk^2+4 \bk \Omega r \CR_2(\alpha)=0 \ , 
\end{equation}
where $\bk \equiv |k| =\sqrt{1-\Omega^2(L^2-r^2)}$ and we introduced the following quantities
\begin{equation} \label{ratios}
\CR_1(\alpha)\equiv \frac{\mathcal T s}{P}=-\frac{4}{1+4\sinh^2\alpha}  \quad \text{and} \quad \CR_2(\alpha)\equiv\frac{Q}{P}=-\frac{4\sinh \alpha \cosh \alpha}{1+4\sinh^2\alpha} \ . 
\end{equation}
We notice that by taking the limit $\alpha\to \infty$ we obtain the EOM \eqref{exl} for the extremal case.

Since the D3-brane is moving on the 5-sphere the local temperature has a redshift as compared to the global temperature $T$ of the background space-time that we are probing as $\mathcal T = T / \bk$. Thus, we are imagining finding the thermal giant graviton solution for a given value of $T$. 
In addition, the above EOM \eqref{exteq} should be supplemented by the charge quantization condition 
\eqref{D3current} 
which from the above becomes
\begin{equation}
\label{charge_con}
N_{\rm D3} T^4 = \frac{2T_{\rm D3}}{\pi^2}  \frac{\sinh \alpha}{\cosh^3\alpha} \bk^4 \ . 
\end{equation}
Our goal is to study the thermal giant graviton for a given value of $N_{\rm D3}$, $T$ and $J$, or, alternatively, $N_{\rm D3}$, $T$ and $r$. From the latter choice we see that $\alpha= \alpha(N_{\rm D3},T,r)$. 

Regarding the validity of our black D3-brane probe in $\ads_5\times S^5$ we see that we need $N_{\rm D3} \gg 1$ to have a valid SUGRA solution of the probe but that at the same time we also need $N_{\rm D3} \ll N$ since the back reaction of the $N_{\rm D3}$ D3-branes of the probe should be negligible in comparison to the back reaction of the $N$ D3-branes from which the $\ads_5\times S^5$ background originates.


\subsection{Solution space \label{sec:thermalsol} }

In this section we examine the structure of the solution to the EOM \eqref{exteq} and the charge quantization constraint \eqref{charge_con}. 

The aim in the following is to study the thermal giant graviton solution given a temperature $T$, the number of D3-branes $N_{\rm D3}$ for the brane probe as well as the radius $r$ of the 3-sphere that the D3-brane probe is wrapped on.%
\footnote{One can alternatively use \eqref{enes} to write $r=r(N_{\rm D3},T,J)$ and consider $N_{\rm D3}$, $T$ and $J$ given, which will be further examined below.}
We find the following two branches of solutions $\Omega_\pm(r)$ to the EOMs \eqref{exteq}
\begin{equation}
\label{omega_pm}
\Omega_\pm(r)=\frac{3}{\sqrt{9L^2-8\big(1+\Delta_\pm(\alpha)\big)r^2}} \ , 
\end{equation}
which we refer to as the lower ($-$) and upper ($+$) branch respectively. 
Here we have defined%
\footnote{Note that $\Omega\to-\Omega$, $\alpha\to-\alpha$ is also a solution. Here we only consider $\Omega>0$ ($\Rightarrow$ $\alpha>0$).}
\begin{equation}
\label{delta_pm}
\Delta_{\pm}(\alpha)=-\frac{1}{8}\left(3\mathcal{R}_1(\alpha)+8\mathcal{R}_2(\alpha)^2\pm 4\mathcal{R}_2(\alpha)\sqrt{\CD(\alpha)} \right)+\frac{1}{2} \spa 
\mathcal D(\alpha) =3\mathcal{R}_1(\alpha)
+4\mathcal{R}_2(\alpha)^2-3 \ , 
\end{equation}
where $\mathcal{R}_{1,2} (\alpha)$ are given in \eqref{ratios}. 
These solutions are supplemented with the charge quantization constraint \eqref{charge_con}. In order to determine $\alpha_\pm$ as a function of $T$, $N_{\rm D3}$ and $r$ one should substitute \eqref{omega_pm}-\eqref{delta_pm} into \eqref{charge_con}.

Taking the extremal limit $\alpha \rightarrow \infty$ we see that $\mathcal R_1 \rightarrow 0$ and $\mathcal R_2 \rightarrow -1$. Hence the above solutions \eqref{omega_pm}-\eqref{delta_pm} reduce to the extremal values $\bar \Omega_\pm $ given in \eqref{Ommext}, 
found in Sec.~\ref{sec:extsol} from the DBI analysis of the extremal giant graviton solution.

Considering Eq.~\eqref{charge_con} we have that $\sinh \alpha / \cosh^3 \alpha$ is bounded from above with maximal value $2\sqrt{3}/9$ corresponding to the value $\bar{\alpha}$ for which $\cosh^2 \bar{\alpha} = 3/2$. Thus, since we choose to be on the branch connected to the extremal D3-brane we always have $\alpha \geq \bar{\alpha}$. Setting $\Omega=0$ and using this bound in Eq.~\eqref{charge_con} gives rise to the maximal temperature for $N_{\rm D3}$ coincident black D3-branes given by $T_{\rm static}= (\frac{4\sqrt{3}}{9\pi^2} \frac{T_{\rm D3}}{N_{\rm D3}})^{1/4}$.
However in the case under investigation  - where $\Omega > 0$ and depends on $\alpha$ - one obtains a stronger bound on $\alpha$ from the requirement that $\mathcal D(\alpha)$ in \eqref{delta_pm} should be always non negative. Indeed, $\mathcal D(\alpha)=0$ for $\alpha = \tilde{\alpha} \equiv \cosh^{-1}(3/2)$ and we see that $\tilde{\alpha} > \bar{\alpha}$. Thus for the finite-temperature giant graviton we have the bound $\alpha \geq \tilde{\alpha}$. 
This means that from \eqref{charge_con} we have the bound
\begin{equation}
\label{klower}
\bk  \geq \hat T \ , 
\end{equation}
where we have introduced the rescaled temperature 
\begin{equation}
\label{Thatrange} 
\hat T \equiv \frac{T}{T_{\rm max}}  \spa T_{\rm max}\equiv  \left( \frac{8\sqrt{5}}{27\pi^2} \frac{T_{\rm D3}}{N_{\rm D3}} \right)^{1/4} \ . 
\end{equation}
Moreover we have the geometric upper bound $r \leq L$ on the size
of the giant graviton, which means that $\bk \leq 1$. It thus follows that for the giant graviton on $S^5$ the value of $\mathbf{k}$ must lie in the range
\begin{equation} 
\hat T \leq \mathbf{k} \leq 1
\end{equation} 
and as a corrollary we see that the temperature is bounded, $\hat T \leq 1$. 
In particular, when the upper bound on the temperature $T_{\rm max}$ is reached the solution space collapses to a point. 
This upper bound on the temperature is more restrictive than for a static black D3-brane where the maximal temperature is $T_{\rm static}$ given above. 

\subsubsection*{Parameterization of thermal giant graviton solution}

We now describe a very useful analytic parameterization of the solution, which will be employed in this and the next section
 to analyze the solution in more detail.  For given $T$ and $N_{\rm D3}$ the solution can be parameterized by the value of  
$\bk$,  as follows.  Introducing
\begin{equation}
\label{phidef}
\phi \equiv \frac{1}{\cosh^2 \alpha}
\end{equation}
one finds that Eq.~\eqref{charge_con} can be rewritten as the cubic equation
\begin{equation}
\label{cubic}
\phi^3 - \phi^2 +  \frac{4}{27} \sin^2 \delta =0     \ , 
\end{equation} 
where we have defined
\begin{equation}
\label{cosdelta}
\sin \delta (\hat T, \mathbf{k})  = \left( \frac{T_{\rm max}}{T_{\rm static}} \frac{\hat T}{\bk} \right)^4  = 
\frac{2 \sqrt{5} }{ 3 \sqrt{3}}  \left(\frac{\hat{T}}{\mathbf{k}}\right)^4 \ . 
\end{equation}
Eq.~\eqref{cubic} is of the same type as encountered for the thermal BIon \cite{Grignani:2010xm,Grignani:2011mr}
and the solution connected to the extremal giant graviton is 
\begin{equation}
\label{alphasol}
\phi (\hat T,\bk ) =   \frac{2}{3} \frac{\sin \delta}{   \sqrt{3}\cos\frac{ \delta}{3} -   \sin \frac{\delta}{3}} \ . 
\end{equation}
We thus have an explicit functional expression for $\alpha (\hat T,\bk)  =  {\rm arccosh} \, \phi (\hat T,\bk)^{-1/2} $
and  substituting this in \eqref{delta_pm} we then obtain $ \Delta_\pm (\hat T,\mathbf{k})$. 
With those expression in hand, one can now obtain $r$ as a function of $(\hat T, \mathbf{k})$, 
using $\mathbf{k}= \sqrt{1- \Omega^2( L^2-r^2})$ and the solution \eqref{omega_pm}. This yields
\begin{equation}
\label{rpm}
r_{\pm} (\hat T, \mathbf{k})  =   \frac{3 \bk} {\sqrt{ 8 \bk^2 (1 + \Delta_\pm (\alpha) ) + 1 - 8 \Delta_\pm (\alpha) }} L  \ , 
\end{equation}
along with
\begin{equation}
\label{Opm}
\Omega_{\pm} (\hat T, \mathbf{k})  = \sqrt{ \frac{ 8 \bk^2 (1+ \Delta_\pm (\alpha) ) + 1 - 8 \Delta_{\pm} }{ 1- 8 \Delta_\pm (\alpha) } }
\frac{1}{L} \ , 
 \end{equation}
 where we recall the range \eqref{Thatrange} for $\bk$.

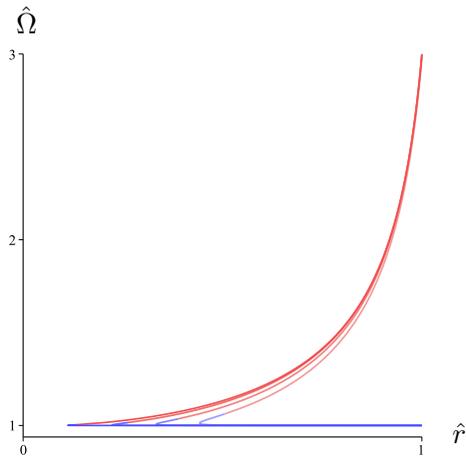
\begin{figure}[!ht]
\centerline{\includegraphics[scale=0.3]{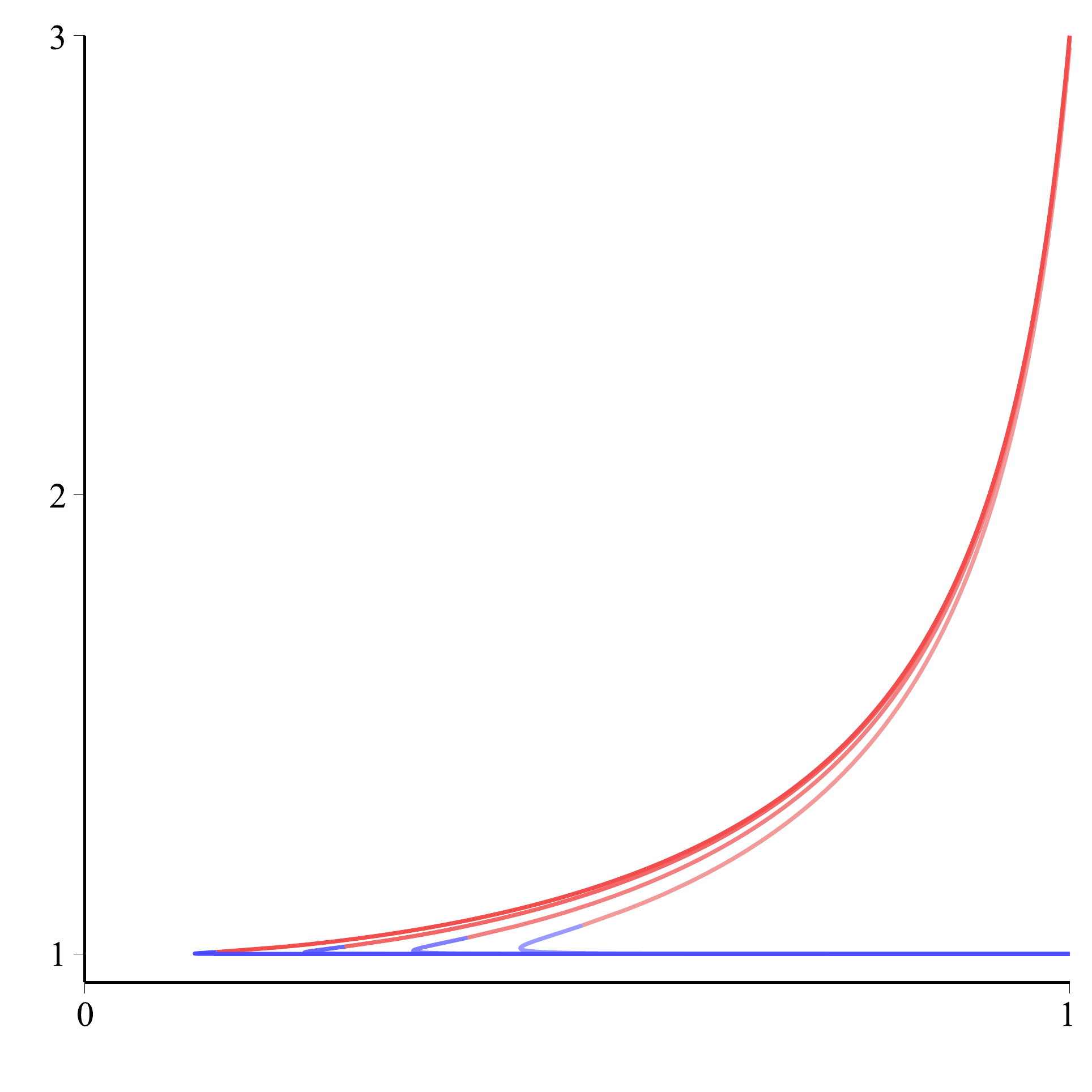}}
\begin{picture}(0,0)(0,0)
\put(310,30){ $\hat r $}
\put(145,185){ $ \hat \Omega   $}
\end{picture}		
\vskip -.5cm
	\caption{$\hat \Omega = \Omega L $ versus  $\hat r = r/L$ for the 
	 two solution branches of thermal giant gravitons for various values of $\hat T$. 
	The lower ($-$) branch is blue and the upper ($+$) branch  is red. 
	Shown are the values $\hat T$ = 0.1, 0.2, 0.3 and  0.4 with temperature increasing
	as the curves move to the right and become fainter in color.} 
	\label{fig:Omegavsr}
\end{figure}

Fig.~\ref{fig:Omegavsr} depicts the resulting solution branches for various values of $\hat T$ in a $(\hat r \equiv r/L, \hat \Omega \equiv \Omega L )$ diagram, and we recall for comparison that the corresponding extremal solution is plotted in
 Fig.~\ref{fig:Omegavsrext}.  We note the following  new and interesting features. 
The lower and upper branch meet in the point where $\bk = \hat T$  
(i.e. $\alpha = \tilde{\alpha}$)  saturating the lower bound in \eqref{klower}. 
At this point we find $\Delta_\pm = - \frac{1}{2}$ giving 
$\Omega_\pm = \tilde{\Omega} = 3 / \sqrt{9L^2 - 4 \tilde{r}^2}$. Inserting this into the expression for $\bk$ in 
\eqref{kexpression} then gives 
\begin{equation} 
\label{rtilde}
\tilde r=\frac{3L}{\sqrt{4+5 \hat T^{-2}} } \ . 
 \end{equation}  
At the upper bound $\bk =1$, we have $r = L$ and $\Omega_- (\hat T, 1) = 1/L$ and
$\Omega_+ (\hat T, 1) \leq 3/L$. 
Furthermore, we observe that the values of $r$ are restricted to $  0 \leq r_{\rm min}( \hat T) \leq r \leq L$, 
and  $r_{\rm min} (\hat T)$ approaches $L$ as the maximum temperature is approached. 
The minimal size thermal giant graviton $r_{\rm min}$ lies on
the lower branch, which curves back to meet the upper branch in the point $\tilde r$. 
Furthermore, for each $r$ value in this range there are two possible
values for $\Omega$, lying in between the two corresponding values of the extremal solution.

\vskip .7cm

That the minimal size of the giant graviton is greater than zero is an important consequence of the finite-temperature physics of the giant graviton. For the extremal giant graviton the two branches meet in the singular solution 
$r=0$ which in turn corresponds to the graviton particle with same angular momentum. 
What we see at finite temperature is that: a) there is a minimal possible size $r_{\rm min}$ of the giant graviton
and b) unlike  in the extremal case, it is possible to move in the solution space from one branch to another
 since the meeting point of the two branches at $\tilde{r}$ is not a singular solution. 
  Note that the fact that the thermal giant graviton attains a minimal possible size has an analogue in the thermal BIon as well as in the thermal Wilson line cases studied in~\cite{Grignani:2010xm,Grignani:2011mr,Grignani:2012iw}.

In the next section we will use the parameterization above to make
a detailed analysis of the thermodynamic quantities of the two thermal giant graviton branches
and examine their stability. In particular, the explicit  parametric solution enables to easily
plot any combination of thermodynamic quantities and study their behavior. 
However, the expressions are too complicated in general to invert but there are  three interesting situations 
which will be considered in more detail. 

\begin{itemize}
\item \textbf{The low-temperature regime:} We can expand around the extremal case by doing an expansion for 
small $\hat T$.  
\item \textbf{The maximal size regime:}  We can expand around $\bk=1$ (i.e. $r = L$). This also includes the maximal 
temperature regime  $T \rightarrow T_{\rm max}$ since in that limit the physics is captured by the large giant graviton limit $\epsilon = (L-r)/L\ll 1$.
\item  \textbf{The minimal charge parameter limit:}  We can expand around $\bk=\hat T$, i.e. 
the point  $\alpha=\tilde\alpha$  where the two branches meet at the value $\tilde r$.  
\end{itemize} 

The first regime above, which is physically the most relevant one, will be investigated further in Sec.~\ref{sec:GGstab},
while the other two limits are considered in App.~\ref{app:points}. 
Since our main focus in the text will be the low-temperature or near-extremal regime,
we now present the solution described above in this limit.

\subsubsection*{Low-temperature solution}

The convenient feature of the parameterization above, using $\phi (\alpha)$ defined in \eqref{phidef}, 
and $\delta$ defined in \eqref{cosdelta},  is that the extremal limit  is obtained for $\phi\to 0$ and
 $\delta \to 0$.  Expanding around this, we can  work out the form of the angular velocity $\Omega$ and charge parameter
  $\phi$ as a function of the thermal giant graviton size $r$ in the low temperature limit.
  
In order to do a low temperature expansion we demand that $\hat{T} \ll 1$ and $\sin \delta\ll 1$.
As a consequence, we need that  $\mathbf{k}\gg \hat{T}$ in order for the expansion to be valid, which in turn implies that  $r\gg \hat{T}L$.
We then have  to leading order in $\hat T$, 
\begin{equation}
\delta \simeq \frac{2\sqrt{5}}{3\sqrt{3}}\left(\frac{\hat{T}}{\mathbf{k}}\right)^4 \spa 
\phi \simeq \frac{2}{3\sqrt{3}}\delta \ , 
\end{equation}
where we used \eqref{cosdelta}, \eqref{alphasol}.  
Performing a small $\delta$ expansion of $\Delta_\pm$ gives then to leading order $\Delta_+\simeq-\phi/4$ and 
$\Delta_- \simeq-1$ (here and below, expressions are up to corrections of order $\hat T^8$).  
This can now be substituted into the expressions \eqref{rpm}  for $r_\pm$, yielding
\begin{equation} 
\label{rpmlowT}
r_- \simeq \mathbf{k}L \spa  r_+ \simeq \frac{\mathbf{k}L}{\sqrt{1+8\mathbf{k}^2}}\left[3+\frac{4\sqrt{5}}{9}\left(\frac{\hat{T}}{\mathbf{k}}\right)^4\frac{ \mathbf{k}^2-1}{1+8\mathbf{k}^2}\right] \ . 
\end{equation}
These expressions can be inverted in order to write $\mathbf{k}$ as a function of $r$. We find 
\begin{equation}
\label{klowT}
\mathbf{k}_- \simeq \frac{r}{L} \spa  \mathbf{k}_+ \simeq \frac{r\bar{\Omega}_+(r)}{3}+\frac{4\sqrt{5}}{r\bar{\Omega}_+(r)}\frac{\rho^2}{r^2}\hat{T}^4 \ , 
\end{equation}
where $\bar \Omega_\pm$ are the extremal values \eqref{Ommext} and where we have defined $\rho^2\equiv r^2-L^2$, while 
the size $r$ now parameterizes the solution and the $\pm$ subscript has thus been moved from $r$ to $\bk$. 

Substituting \eqref{klowT}  into the expression \eqref{Opm} for $\Omega$ we then obtain 
\begin{equation}
\Omega_-(r) \simeq \frac{1}{L} \spa 
\Omega_+(r)\simeq \bar{\Omega}_+(r)\Bigg[1-\frac{4\sqrt{5}}{3 r^2\bar{\Omega}_+^2(r)} \hat{T}^4 \Bigg] \ , 
\end{equation}
along with 
\begin{equation}
\label{philowT}
\phi_-(r) \simeq \frac{4\sqrt{5}}{27}\frac{L^4\hat{T}^4}{r^4} \spa 
\phi_+(r) \simeq 12\sqrt{5}\frac{\hat{T}^4}{r^4\bar{\Omega}_+^4(r)} \ .   
\end{equation}
Note that the angular velocity on the upper branch is much more sensitive to low temperatures than the lower branch.

\subsection{Validity of probe approximation}%

For the probe approximation to be valid for our SUGRA black D3-brane probe we must require the transverse length scale $r_s$ of the probe to satisfy the following conditions
\beq 
\label{valc}
r_s\ll r_{\rm int} \spa r_s\ll r_{\rm ext} \spa r_s\ll L \ , 
\eeq
where $r_{\rm int}$ and $r_{\rm ext}$ are the length scales associated with the intrinsic and extrinsic curvature of the embedding of the brane, respectively, and $L$ is the length scale of the $\ads_5 \times S^5$ background.
For the black D3-brane in the branch connected to the extremal solution the transverse length scale $r_s$ is easily seen to be given by $r_s \sim ( \frac{N_{\rm D3}}{T_{\rm D3}} ) ^{1/4} $ \cite{Grignani:2010xm}.

We compute the Ricci scalar for the embedding metric in order to obtain the intrinsic length scale. This simply gives $r_{\rm int} = r / \sqrt{6}$. Instead the extrinsic length scale of the embedding is obtained as $r_{\rm ext}=|K^{\rho}n_{\rho}|^{-1}$ where $n_\rho$ is the unit normal vector to the brane embedding and $K^\rho$ is the extrinsic curvature. We find
\beq
r_{\rm ext}=\frac{r\bk^2}{(1-\frac{r^{2}}{L^2})}\frac{\sqrt{\bk^2+1-r^2/L^2}}{\Omega^2r^2+3\bk^2} \ . 
\eeq
Collecting now this information with \eqref{valc} we see that we have two different regimes to consider for the validity of the probe approximation, namely whether $r/L$ is small or not (note of course that $0 \leq r/L \leq 1 $). If $r/L$ is not small, then we can roughly regard $r$ and $L$ to be of the same order, hence \eqref{valc} simply reduces to $r_s \ll L$ which using $L^4 = N / ( \Omega_{(3)} T_{\rm D3} )$ one can write as
\begin{equation}
N_{\rm D3} \ll N \ , 
\end{equation}
a condition already mentioned in the beginning of this section. Instead, if $r/L$ is small we should impose the condition that $r_s \ll r$ which we can write as $N_{\rm D3} /N \ll r^4 / L^4$. 

In addition to the validity of the probe approximation we should also require the validity of the SUGRA description of the black D3-brane probe. This requires $N_{\rm D3} \gg 1$ as well as $g_s N_{\rm D3} \gg 1$. The latter condition can be written as $\lambda N_{\rm D3} \gg N$ ($\lambda$ being the 't Hooft coupling). Therefore, if we assume we are in the regime where $r/L$ is not small we can summarize both the probe approximation condition as well as the conditions for validity of the SUGRA D3-brane description as the conditions
\begin{equation}
\label{validity}
1 \ll N_{\rm D3} \ll N \ll \lambda N_{\rm D3} \ . 
\end{equation} 

\subsubsection*{Hawking-Page temperature}

An important point is to examine how the bounds above relate to the Hawking-Page temperature, above
which the $\ads$ black hole background will become dominant over the hot $\ads$ spacetime background
considered in this paper.  Using that  the Hawking-Page temperature $T_{\rm HP} \sim 1/L$, and the expression
for $T_{\rm max}$ in \eqref{Thatrange}, we find
\beq
\frac{T_{\rm max}  }{T_{\rm HP}} \sim \left( \frac{N}{N_{\rm D3}} \right)^{1/4} \gg 1 \ , 
\eeq
where we used \eqref{validity} in the last step. Thus in the regime where the probe blackfold approximation
is valid the maximum temperature that the solution exhibits is far above the Hawking-Page temperature.
Consequently, this maximum temperature is not physical in the sense that before reaching it one should
change the background to the $\ads$ black hole. In particular, this means that our solution should be
considered for small temperatures (much less than $ T_{\rm max}$) only.  
This is very similar  to the case of thermal string probe in $\ads$ considered in Ref.~\cite{Grignani:2012iw}.


\section{Thermodynamics and stability properties \label{sec:GGstab}} 

In this section, we further investigate the thermal giant graviton solution obtained in Sec.~\ref{sec:GGthermal}.
We will first compute the relevant conserved quantities and thermodynamic properties
using the formulae derived in Sec.~\ref{sec:charges}. This will also enable us to show that the same solution can
be derived from an action and verify the first law of thermodynamics. 
Then the solution parameterization of the previous
section will be used to examine the detailed behavior of the thermodynamics
and determine which part of the solution branches are stable.  Furthermore, we make a detailed analysis
of the low temperature regime. 

\subsection{Thermodynamic quantities and first law}

Using the expressions in Eqs.~\eqref{Mres2}, \eqref{BFentropy} together with the perfect
fluid blackfold stress tensor (see \eqref{thermoquantities}, \eqref{TandS}) and current \eqref{D3current}, 
 the (off-shell) energy $E$, angular momentum $J$ and entropy $S$ are computed to be%
\begin{equation}
\label{enes}
E(r)=  \frac{ \Omega_{(3)} \epsilon(r)r^3}{\sqrt{1-\Omega^2 (L^2-r^2)}} \spa 
J(r)=\Omega E(r)(L^2-r^2)+ \Omega_{(3)} Q r^4 \spa 
S(r)= \Omega_{(3)} r^3 s(r) \ , 
\end{equation}
with
\begin{equation}
\label{epsilons}
\epsilon(r)=  \frac{T_{\rm D3}^2}{2}  
\frac{\left(1-\Omega^2 (L^2-r^2)\right)^2}{\pi^2T^4}
\frac{ 5+4\sinh^2\big(\alpha(r)\big) }{  \cosh^4 \big(\alpha(r)\big)} 
\spa
s(r)=2 T_{\rm D3}^2 \frac{\left(1-\Omega^2 (L^2-r^2)\right)^{5/2}}{\pi^2T^5\cosh^4 \big(\alpha(r)\big)} \ , 
\end{equation}
where $\Omega_{(3)} = 2\pi^2$.

We can use these quantities to compute the thermodynamic action \eqref{actE} (see also App.~\ref{app:thermo}),
which is the Gibbs free energy,
\begin{equation}
\label{actggthermal}
 \beta I_E  = F =   E -T S - \Omega J =   -  \Omega_{(3)}  \frac{T_{\rm D3}^2}{2}   (r^3 \bk P +r^4  \Omega Q) \ , 
  \end{equation}
where $P$ is the pressure defined in \eqref{thermoquantities}. 
 Varying with respect to $r$ keeping $T$, $\Omega$ and $Q$ constant, one may check explicitly that  one 
indeed obtains the EOM \eqref{exteq}, $i.e.$ the equation
\begin{equation}
\frac{\text{d}E(r)}{\text{d} r}-\Omega \frac{\text{d}J(r)}{\text{d} r}-T\frac{\text{d}S(r)}{\text{d} r} = 0 \ , 
\end{equation}
is equivalent to the EOM \eqref{exteq}. We see that this in turn means that the first law of thermodynamics is satisfied for solutions of the EOM \eqref{exteq}.

To obtain the on-shell expressions for the conserved quantities in \eqref{enes}, it is useful to define the
rescaled quantities
\begin{equation}
\label{EJSdef}
\mathbf{E} \equiv  \frac{ L E  }{N_{\rm D3} N}   \spa \mathbf{J} \equiv  \frac{ J}{ N_{\rm D3} N }  \spa 
\mathbf{S}  \equiv   \frac{ L  T_{\rm max} S }{N_{\rm D3} N }  \  ,  
\end{equation}
where we recall that $N_{\rm D3} N = \Omega_{(3)} Q L^4$. 
These satisfy the first law of thermodynamics
in the form $d \mathbf{E} = \hat T d \mathbf{S} + \hat{\Omega} d\mathbf{J}$, where $\hat \Omega \equiv \Omega L$. 
One can also check that the Smarr relation \eqref{Smarr} is satisfied. 
In the following we will also use again $\hat r \equiv r/L$ to simplify the equations.  
Using \eqref{enes} and the definitions above, one finds
\begin{equation}
\label{EJonshell}
\mathbf{E}_\pm  (\hat T,\bk) = \frac{27}{16 \sqrt{5}}  \frac{ \bk^3  \hat r_\pm^3  }{\hat T^4} 
\phi ( 4  + \phi) 
\spa 
\mathbf{J}_\pm  (\hat T,\bk) =   \mathbf{E}_\pm  \hat \Omega_\pm  (1- \hat r_\pm^2) + \hat r_\pm^4 \ , 
\end{equation}
\begin{equation}
\label{Sonshell}
\mathbf{S}_\pm    (\hat T,\bk)  = \frac{27}{4 \sqrt{5}}  \frac{   \bk^5  \hat r_\pm^3 }{\hat T^5  } 
\phi^2  \ , 
\end{equation}
where $\phi (\hat T,\bk)$  is determined by \eqref{alphasol}   and $\hat r_\pm (\hat T, \bk)$  for the two solution
branches given in \eqref{rpm}.  It also follows that the on-shell free energy is given by
\begin{equation}
\label{Fred}
\mathbf{F}_\pm  (\hat T,\bk)  = 
\frac{27}{16 \sqrt{5}} \frac{ \bk^5  \hat r_\pm^3  }{\hat T^4} 
 \phi (4 - 3 \phi)  - \hat \Omega \hat r_\pm^4 \ . 
\end{equation} 
These explicit results enable one to easily plot and examine any combination of the above quantities including 
$\hat r_\pm$ for given value of $\hat T$.   

\begin{figure}[!ht]
\centerline{\includegraphics[scale=0.3]{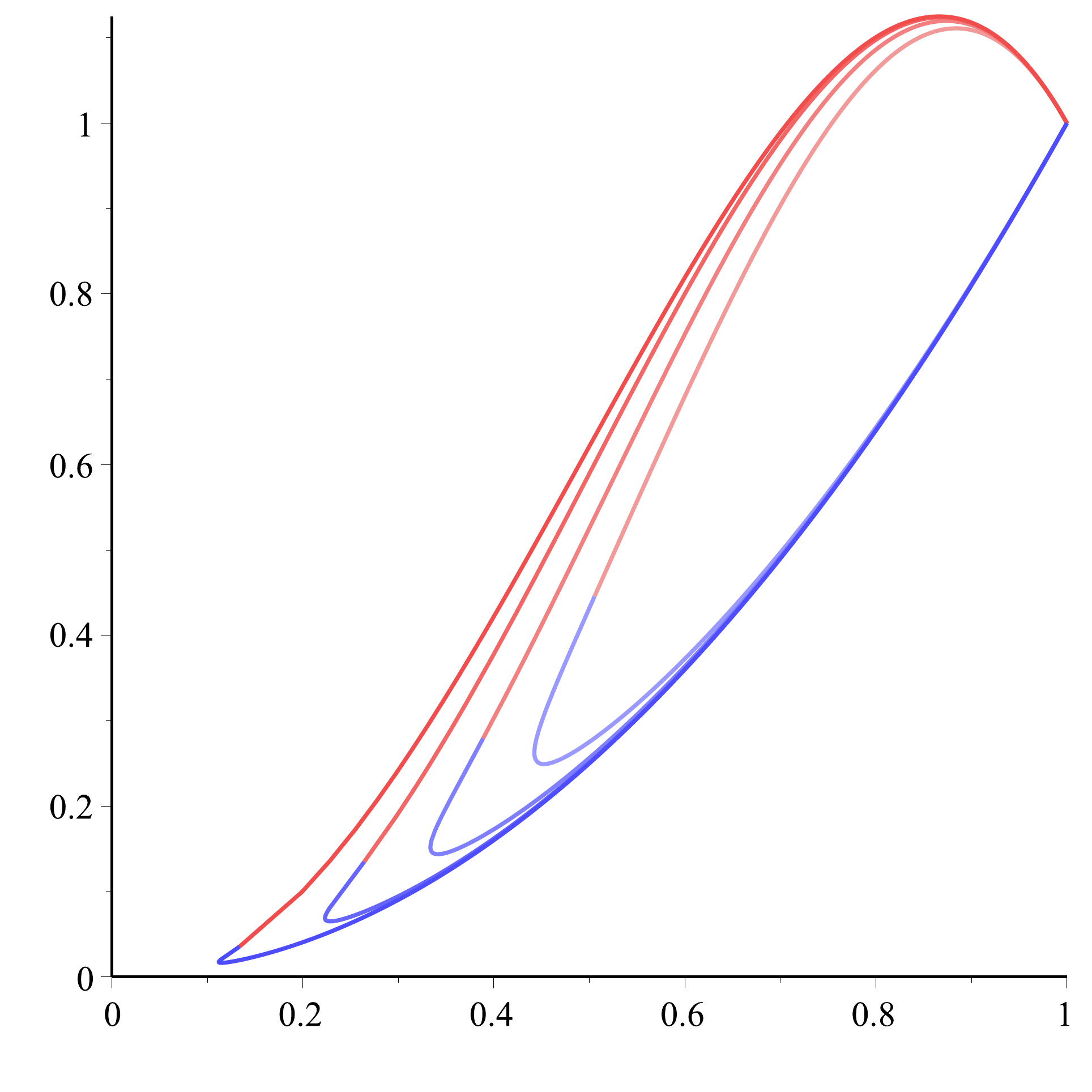}}
\begin{picture}(0,0)(0,0)
\put(310,30){ $\hat r $}
\put(145,185){ $ \mathbf{J}    $}
\end{picture}		
\vskip -.5cm
	\caption{$\mathbf{J}$   versus  $\hat r $ for the 
	 two solution branches of thermal giant gravitons for the same values of $\hat T$
	  as in Fig.~\ref{fig:Omegavsr}. 	} 
	\label{fig:Jvsr}
\end{figure}
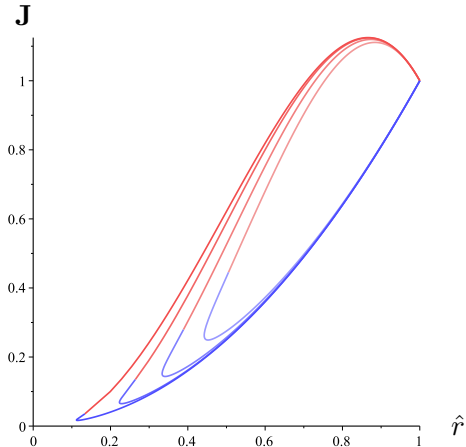

To start, we exhibit in Fig.~\ref{fig:Jvsr} the solution branches in a $(\hat r, \mathbf{J})$ plot for various values of $\hat T$, 
and we recall that the corresponding figure for the extremal case is given in Fig.~\ref{fig:Omegavsrext}. 
We note the following new features. The angular momentum lies in the range $J_{\rm min} (\hat T) \leq J \leq 
J_{\rm max} (\hat T) $, where the boundary values satisfy $d J(r)/dr =0$.  
We will denote the corresponding $r$ values by $r_{J_{\rm min}}$ and $r_{J_{\rm max}}$ respectively. 
Contrary to the extremal case  we observe that one has a non-zero lower bound $J_{\rm min} (\hat T)$ 
which increases with $\hat T$. On the other hand, the upper bound decreases with temperature. 
Just as for the extremal case, we observe that for each $J$ in this range, there are two solutions depending
on the value of $r$. We will shortly see that the one with largest $r$ is the one that is stable.

\vskip .7cm

\subsection{Stability} 

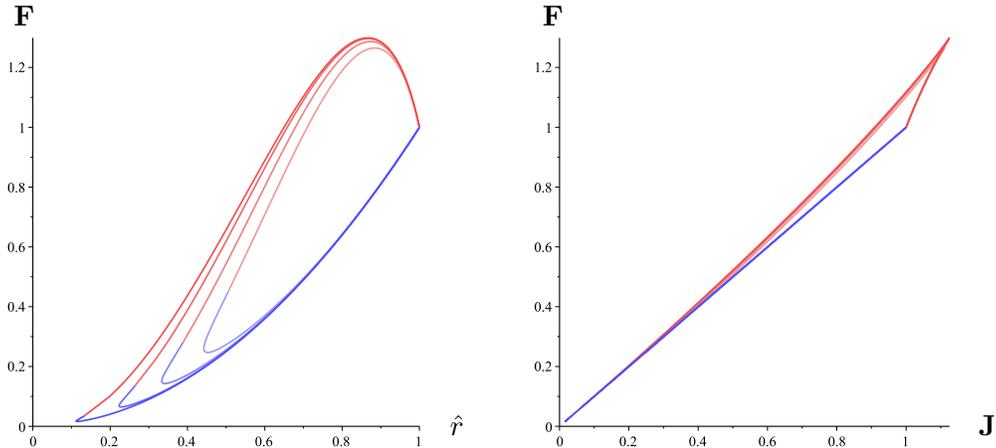
\begin{figure}[!ht]
\centerline{ \includegraphics[scale=0.3]{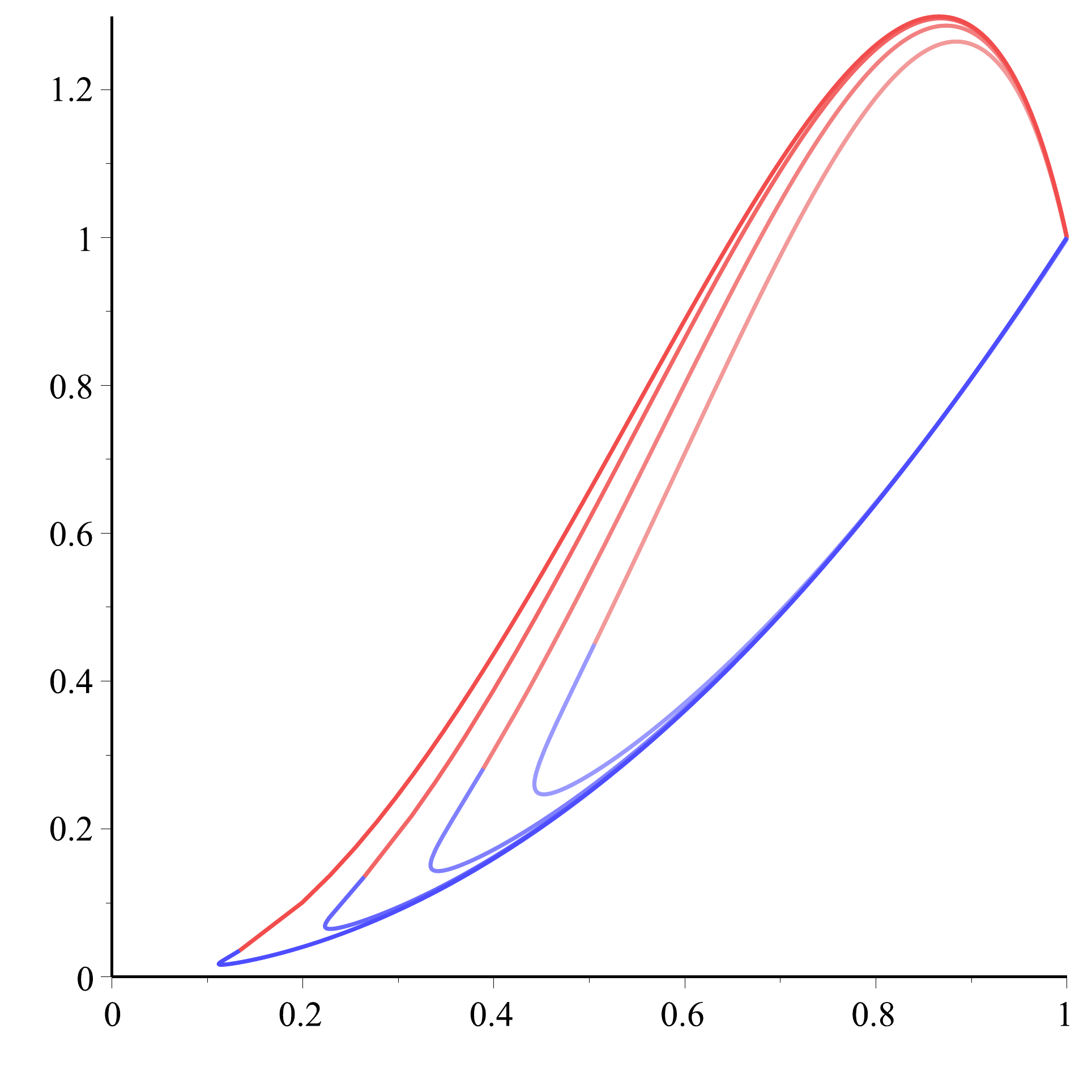} \hskip 1cm
\includegraphics[scale=0.3]{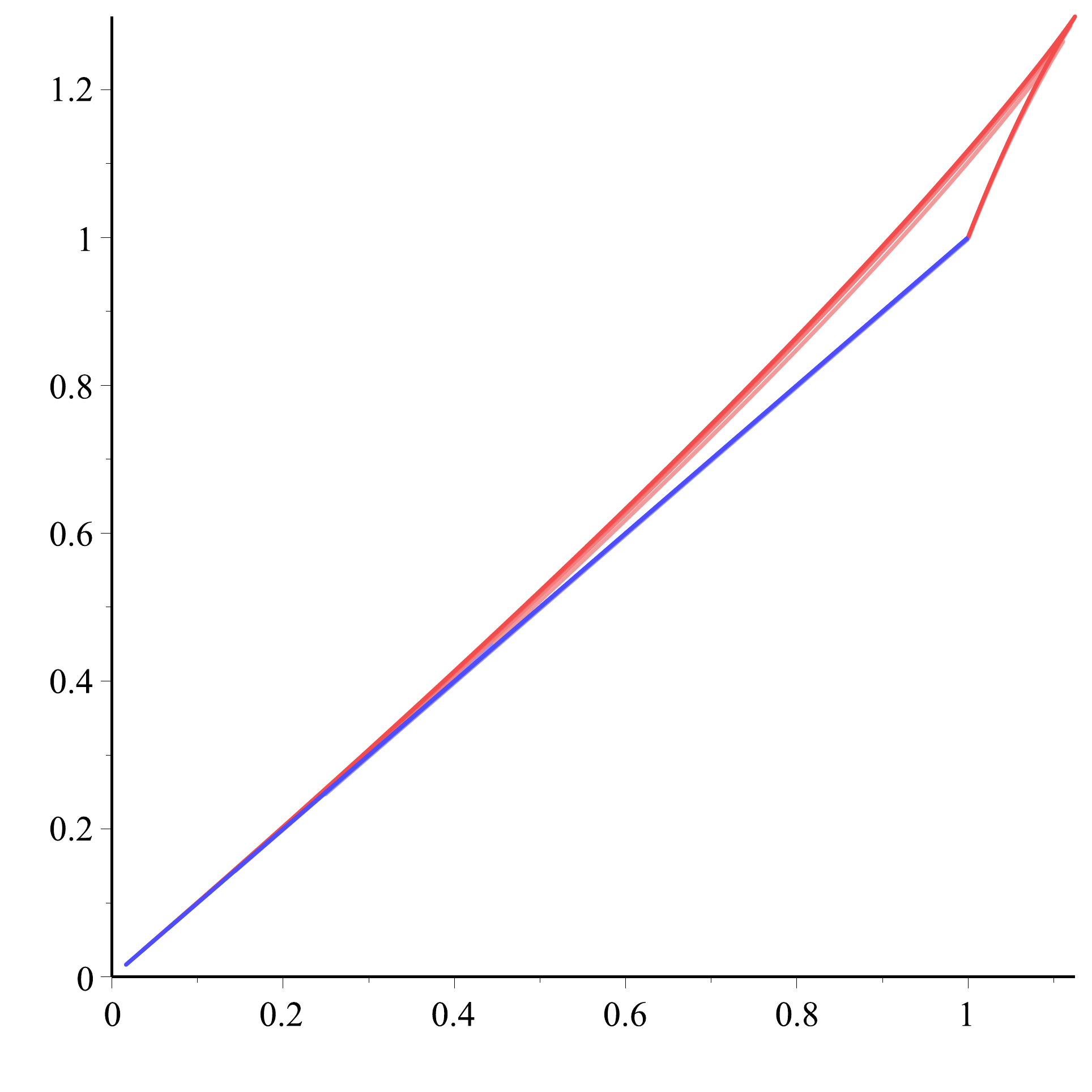} }
\begin{picture}(0,0)(0,0)
\put(210,30){ $\hat r $}
\put(45,185){ $ \mathbf{F}   $}
\put(410,30){ $\mathbf{J} $}
\put(245,185){ $ \mathbf{F} $}
\end{picture}		
\vskip -.5cm
	\caption{$\mathbf{F} $ versus $\hat r$ (left plot) and versus  $\mathbf{J}$ (right plot) for the two solution branches of thermal  giant gravitons at  the same values of $\hat T$
	  as in Fig.~\ref{fig:Omegavsr}. 
	} 
	\label{fig:fvsr}
\end{figure}

To address the stability we turn our attention to the on-shell free energy 
given in \eqref{Fred}.   In Fig.~\ref{fig:fvsr} we have depicted $ (\hat r, \mathbf{F})$
  as well as    $(\mathbf{J},\mathbf{F})$ plots for various values of $\hat T$. Note that the corresponding
  plots for the energy in the extremal case were given in Fig.~\ref{fig:Evsrext}. 
Comparison of the free energies then shows that the lower branch is expected to be stable for
$  r_{J_{\rm min}} \leq r \leq L$ (with $\mathbf{J}_{\rm min}  \leq \mathbf{J} \leq 1$)
and the upper branch for $ r_{J_{\rm max}} \leq r \leq 1$ 
(with $   1 \leq \mathbf{J} \leq \mathbf{J}_{\rm max}$). This is entirely in parallel with the
stability properties of the extremal giant graviton (see Sec.~\ref{sec:extsol}), the difference being
that as a consequence of the finite temperature, a part of the lower branch has become unstable
 and there is a minimum angular momentum. Note that it follows that the minimum size {\it stable}
 thermal giant graviton is thus $r_{J_{\rm min}}$, which is greater than $r_{\rm min}$ (for which
 the solution is unstable).  We also see that the point where the branches meet in $\tilde r$
 is always in the unstable region. On the other hand, the branches also meet in $r= L$,
 but for different values of $\Omega$.  These special points will be considered in more detail
 in App.~\ref{app:points}.

\vskip .7cm

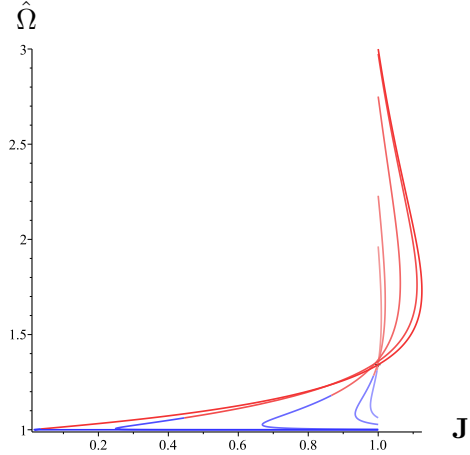
\begin{figure}[!ht]
\centerline{\includegraphics[scale=0.3]{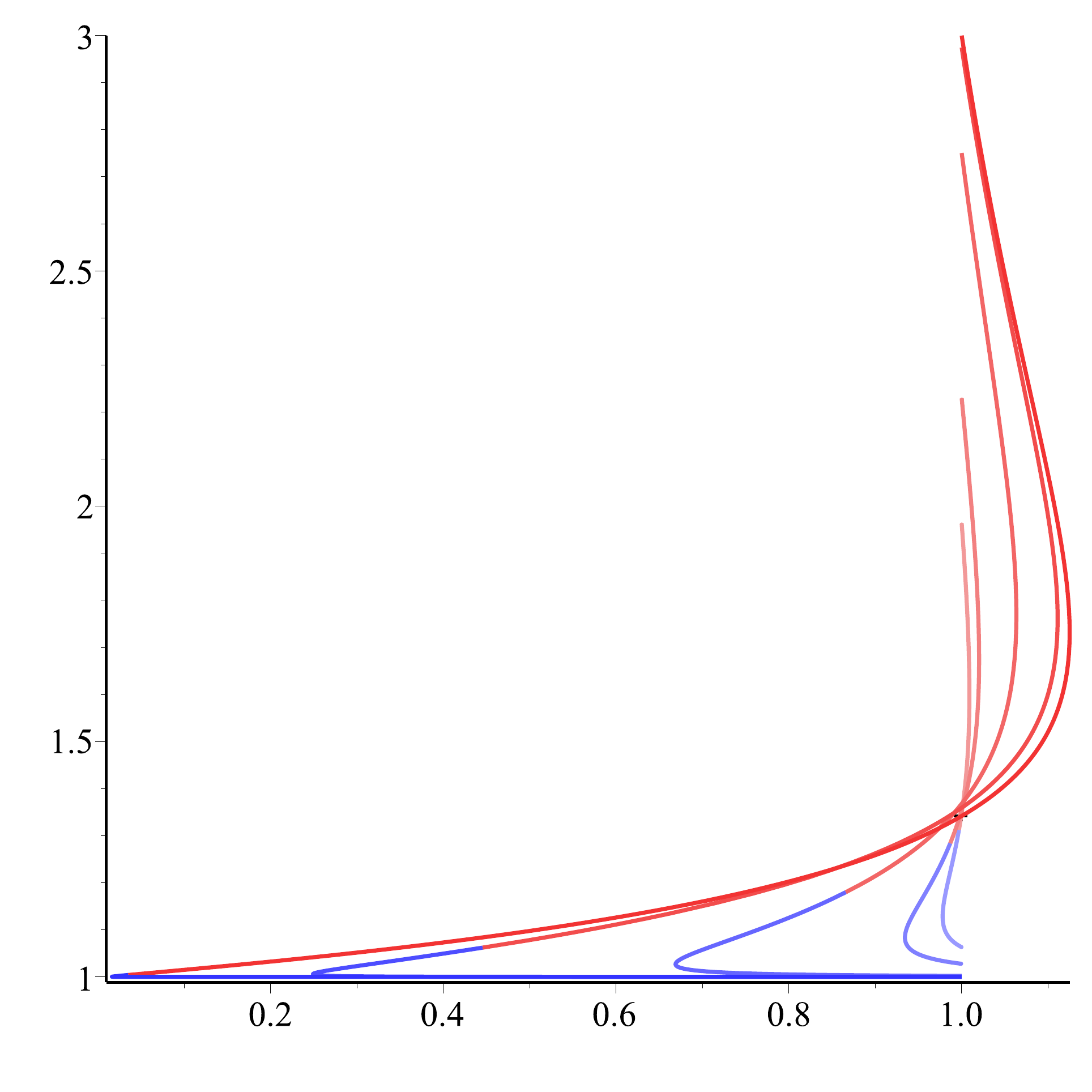}}
\begin{picture}(0,0)(0,0)
\put(310,30){ $ \mathbf{J}   $}
\put(145,185){  $\hat \Omega $}
\end{picture}		
\vskip -.5cm
	\caption{$\hat \Omega$ versus $\mathbf{J}$   for the 
	 two solution branches of thermal giant gravitons for the same values of $\hat T$
	  as in Fig.~\ref{fig:Omegavsr}. 	} 
	\label{fig:Omegavsj}
\end{figure}

The fact that $J_{\rm min}$ and $J_{\rm max}$ denote the onset of instability in the lower and upper
branch respectively is further corroborated by looking at the turning points in a 
$(\mathbf{J},\hat \Omega)$ plot, which is shown  in Fig.~\ref{fig:Omegavsj}.  
We see that these boundaries of stability occur precisely at the turning points where $dJ/d \Omega = 0$,
in accord with expectations based on the Poincar\'e turning point method (see e.g. 
\cite{Arcioni:2004ww} and references therein). 
Finally, we note that these results for the stability of the branches are confirmed by a more detailed off-shell analysis
for the three limits described in Sec.~\ref{sec:thermalsol}. The special meeting points  of the two branches
$\bk = 1$ and $\bk =\hat T$, which correspond respectively to the maximum size
thermal giant graviton and the minimal charge parameter solution and their stability are considered in 
App.~\ref{app:points}.  The most  interesting limit, which is the low temperature limit, will be
considered in the next subsection.

\vskip .7cm

\subsection{Low-temperature limit \label{sec:lowTlimit} }  

We now compute the various thermodynamic quantities for the lower and upper solution branch for low temperatures. 
Using the expansions \eqref{klowT}, \eqref{philowT} in  \eqref{EJonshell}, \eqref{Sonshell},
one finds for the energy, angular momentum and entropy for the lower branch in the low temperature limit 
\begin{equation}
\label{Elower}
\begin{split}
\mathbf{E}_- (\hat T, \hat r) \simeq  \hat{r}^2+\frac{\sqrt{5}}{9}\frac{1}{\hat r^2}\hat T^4 \spa 
\mathbf{J}_- (\hat T,\hat r) \simeq \hat{r}^2+\frac{\sqrt{5}}{9}\frac{\hat \rho^2}{\hat r^2}\hat T^4 \spa 
\hat{T}\mathbf{S}_- (\hat T,\hat r) \simeq \frac{4\sqrt{5}}{27}\hat T^4 \ , 
\end{split}
\end{equation}
where here and in the following $\hat \rho^2 = 1- \hat r^2$.  
The free energy becomes
\begin{equation}
\label{Flower} 
\mathbf{F}_- (\hat T, \hat r) \simeq \hat{r}^2+\frac{\sqrt{5}}{9\hat r^2}\left(1-\frac{4\hat r^2}{3}\right)\hat T^4 \ , 
\end{equation}
and we recall that the higher order corrections in the expressions above are of order $\hat T^8$.
It is trivial to check that the first law $ \text{d}\mathbf{E}_-=\hat \Omega\text{d}\mathbf{J}_- +\hat T\text{d}\mathbf{S}_-$ is satisfied for these quantities. 

We can also compute the 
on-shell free energy as a function of the lower branch angular momentum. 
To this end we invert $\mathbf{J}$ in \eqref{Elower}
\begin{equation}
\label{rminusJ}
\hat r_- (\hat T, \mathbf{J}) \simeq \sqrt{\mathbf{J}} + \frac{3 \sqrt{5} }{40} \frac{ \mathbf{J}-1} {  \mathbf{J}^{3/2}}
\hat T^4  \ . 
\end{equation}
Inserting this into \eqref{Flower} we then obtain 
\begin{equation}
\label{FlowerlowT}
\mathbf{F}_-(\hat T, \mathbf{J} )= \mathbf{J} - \frac{ \sqrt 5}{27}\hat T^4  +{\cal{O}} ( \hat T^8)  \ , 
\end{equation}
where we note that the first correction to the extremal results is independent of the angular momentum. 
Reintroducing the dimensions from the definitions \eqref{EJSdef}, this gives us the final result
for the low temperature expansion of the free energy of the lower branch thermal giant graviton 
\beq
\label{FlowerlowT2}
F (T,J) = \frac{J}{L} - \frac{\pi^4}{4}  N_{\rm D3}^2 L^3 T^4    + {\cal{O}}(T^8) \ . 
\eeq
This is one of the central results of the paper. 

 Finally we can compute the ratio $J/E$ for the lower branch. We find 
\begin{equation}
\frac{J}{E}  = L  -\frac{3\pi^4 L}{4J}N_{\rm D3}^2 (LT)^4  + 
{\cal{O}} (T^8)    \  , 
\end{equation}
where the first term is recognized as the usual Kaluza-Klein contribution while the second term is due to thermal effects. 

Repeating this procedure for the upper branch, we find
\begin{equation}
\label{Eupper}
\mathbf E_+ (\hat T, \hat r) \simeq \frac{3\hat r^2}{\hat{\bar{\Omega}}_+}\left[1+\frac{\sqrt{5}(\hat r - 3\hat  \rho^2)}{\hat r^4\hat{\bar{\Omega}}_+^2}\hat T^4\right] \ , 
\end{equation}
\begin{equation}
\label{JpluslowT}
 \mathbf J_+ (\hat T, \hat r) \simeq
\hat r^2(3\hat \rho^2+\hat r^2)\left[1-\frac{9\sqrt{5} \hat \rho^2}{\hat r^4(3\hat \rho^2+\hat r^2)\hat{\bar{\Omega}}_+^4}\hat T^4\right]\spa 
\hat T \mathbf S_+ (\hat T, \hat r) \simeq\frac{4 \sqrt{5}}{\hat{\bar{\Omega}}_+^3}\hat T^4 \ . 
\end{equation}
Again it is possible to show that these quantities obey the first law of thermodynamics.
Finally, for the free energy one finds in this case
\begin{equation}
\mathbf F_+ (\hat T , \hat r) =\frac{3\hat r^2}{\hat{\bar{\Omega}}_+}
\left(1-\frac{3\sqrt{5}}{\hat{\bar{\Omega}}_+^4 \hat r^4}\right) \ . 
\end{equation}
One may eliminate from this $\hat r$ in favor of $\mathbf{J}$ as done above for the lower branch, but
the resulting expression involves a complicated function of the angular momentum in multiplying the
thermal correction, so that we omit it here. 

We note that the results for $\mathbf{J}$ in  \eqref{Elower}, \eqref{Eupper}  
explicitly show what is seen in Fig.~\ref{fig:Omegavsr}, namely that turning on a temperature has the
 effect that the lower ($-$) branch is pushed to the right while the upper ($+$) branch  is pushed to the 
left. 

\subsubsection*{Stability} 

We now turn our attention to stability.  The method we use is described in Sec.~\ref{stab_method} and is based on an analysis of the Helmholtz free energy $F_{\text{H}}\equiv E-TS$ whose on-shell (rescaled) value is related to the Gibbs free energy \eqref{Fred} through the relation
\beq
\textbf{F}_{\text{H}}^{\pm}=\textbf{F}_{\pm}+\hat{\Omega}\hat{r}^4 \ . 
\eeq
Having described the solution space for low temperatures means that we have essentially solved 
the first derivative $(\mathbf{F}_{\text{H}})_{(1)}$ of the off-shell free energy  for $\hat T\ll 1$, $r\gg \hat T L$.
To analyze the stability, we thus compute the second derivative $ (\mathbf{F}_H)_{(2)}$ for both 
 branches. We find to leading order in $\hat T$
\begin{equation}
(\mathbf{F}_H)_{(2)}^- \simeq \frac{2 \hat r^2}{\hat \rho^2}\left[ 1 - \frac{\sqrt 5 (7-4\hat r^2)}{9\hat r^4}\hat T^4 \right]  \ , 
\end{equation}
and
\begin{equation}
(\mathbf{F}_H)_{(2)}^+ \simeq \frac{2\hat{\bar{\Omega}}_+ \hat r^2 }{3  \hat \rho^2}\left[
\left(4 \hat r^4-3\right)-\frac{\sqrt{5}}{\hat{\bar{\Omega}}_+^2\hat r^4}\left(27-40 \hat r^2+16 \hat r^4\right)\hat{T}^4\right] \ . 
\end{equation}
Solving for $(\mathbf{F}_H)_{(2)}=0$ determines where a solution goes from stable to unstable. Since the low temperature expansion is only valid for $r\gg \hat T L$, we see that the entire part of the lower branch captured by the low temperature expansion remains stable. However,
the value of $r$ for which the upper branch becomes unstable is pushed up when we turn on a temperature. Indeed, solving $ (\mathbf{F}_H)_{(2)}^+=0$, we find that the upper branch becomes unstable at 
\begin{equation}
r_* = \frac{\sqrt{3}}{2}L+\frac{8L}{9}\sqrt{\frac{5}{3}}\hat T^4 + {\cal{O}} (\hat T^8) \ . 
\end{equation}
Here the first term is recognized as the zero temperature instability from the DBI analysis
in Sec.~\ref{sec:extsol}.  Note also that as a consistency check, the same value of $r_*$ is obtained
by finding the maximum of $\mathbf{J}_+$ in the low-temperature expansion \eqref{JpluslowT},
i.e. $ (\partial \mathbf{J}_+/ (\partial r) \vert_{r=r_*} =0 + {\cal{O}} (\hat T^8) $.

\subsubsection*{Maximal and minimal angular momentum}

We also derive the low temperature limit expression for the maximal and minimal value of the 
angular momentum, found on the upper and lower branch respectively. 
 The largest value of $\mathbf J$ is exactly attained on the upper branch where it goes from stable to unstable. So $\mathbf J_\text{max}=\mathbf J_+(r_*)$. Using this, we 
find 
\begin{equation}
\mathbf J_\text{max}=\frac{9}{8}-\frac{\sqrt{5}}{3}\hat{T}^4 + {\cal{O}} (\hat T^8) \ . 
\end{equation}
This expression fits nicely with the numerical data. 

\begin{figure}[!ht]
\centerline{\includegraphics[scale=0.3]{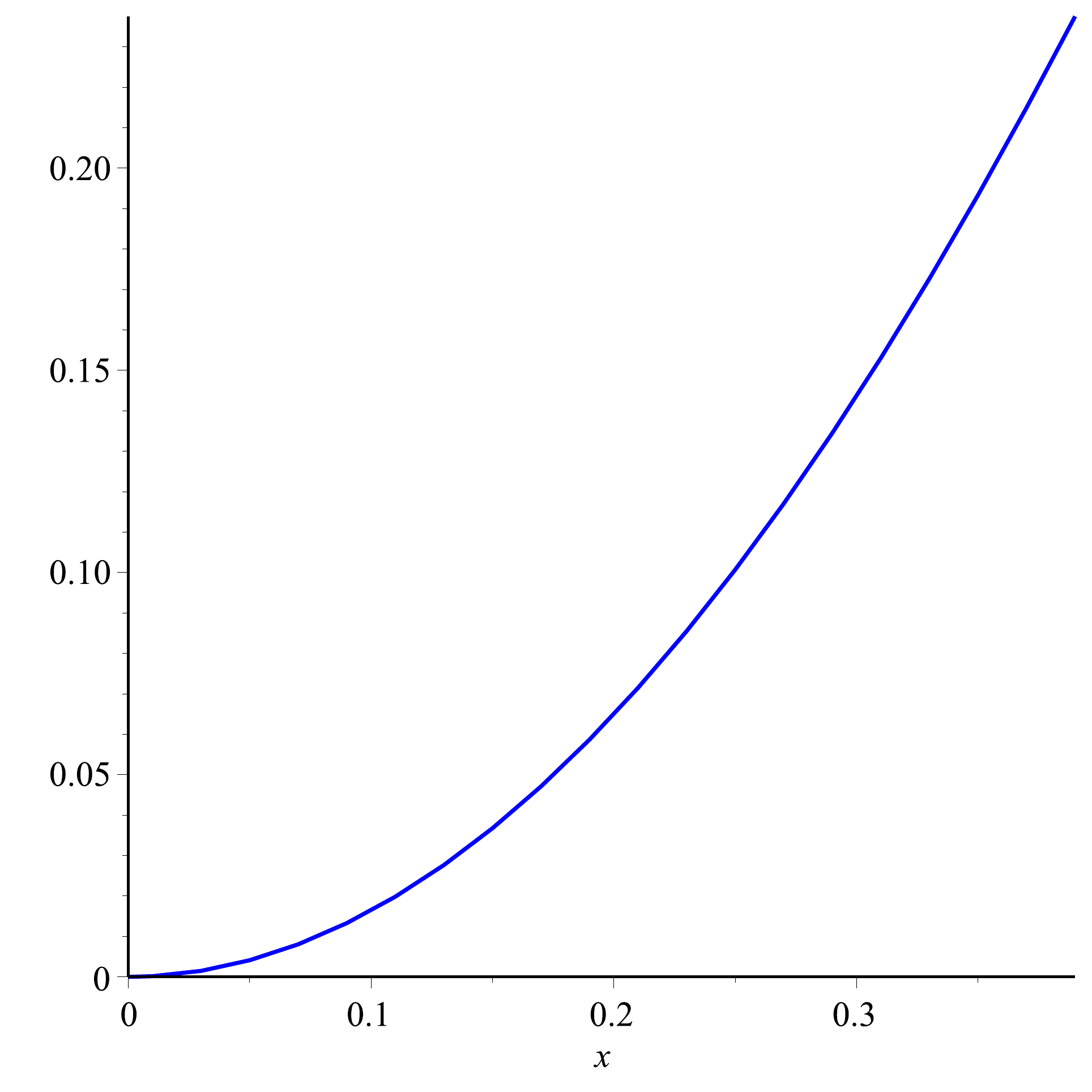}}
\begin{picture}(0,0)(0,0)
\put(310,30){$ \hat T  $ }
\put(145,185){ $\mathbf J_\text{min} $ }
\end{picture}		
\vskip -.5cm
	\caption{The minimum angular momentum  $\mathbf J_\text{min} $  of the thermal giant graviton versus the temperature
	$\hat T $.} 
	\label{fig:JminvsT}
\end{figure}
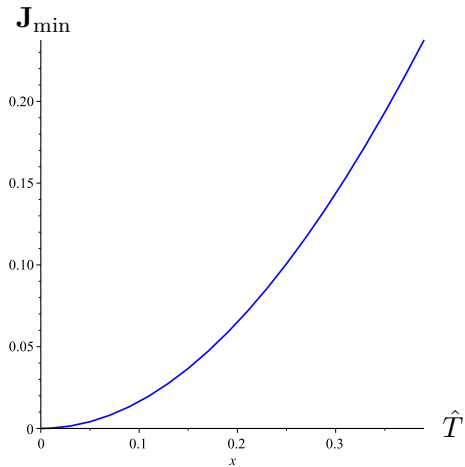

The minimal value of $\mathbf J$ is attained close to $r=0$. This means that an analytical expression for $\mathbf J_\text{min}$ is not obtainable from the low temperature expansion (as it is only valid for 
$\hat r\gg \hat T$). However, we expect the following behavior for small $\hat{T}$
\begin{equation}
\mathbf J_\text{min} \sim \hat{T}^\beta \ . 
\end{equation}
It is then possible to do a fit of the numerically obtained values for $J_\text{min}$. Doing this one finds that $\beta \approx 1.89$. A plot of numerical values of $\mathbf J_\text{min}$ versus $\hat T$ is given in Fig.~\ref{fig:JminvsT}.

\vskip .7cm

\section{Thermal giant graviton on $\ads_5$ \label{sec:GGads} } 

In this section we analyze the case of thermal giant gravitons moving in the $\ads_5$ part of the $\ads_5 \times S^5$ background.
We briefly review the extremal configuration of Refs.~\cite{Grisaru:2000zn,Hashimoto:2000zp} and then present its thermal generalization following the prescriptions used in Secs.~\ref{sec:GGthermal} and \ref{sec:GGstab} for the case of 
thermal giant gravitons moving on the $S^5$ part.

\subsection{Extremal giant graviton on $\ads_5$} 

The $\ads_5$ metric is parameterized as
\beq
\label{adstimess}
ds^2=-\left(1+\frac{\rho^2}{L^2}\right)dt^2+\left(1+\frac{\rho^2}{L^2}\right)^{-1}d\rho^2+\rho^2d\Omega_{(3)}^2 \ , 
\eeq
with the four-form RR gauge field on $AdS_{5}$ given by
\beq
\label{ads5gauge}
A_{t\alpha_1...\alpha_{3}}=-\frac{r^4}{L}d  \Omega_{(3)} \ , 
\eeq
where the coordinates  $\alpha_i$ parametrize the $\Omega_{(3)}$ sphere in $\ads_5$. 
We use the embedding 
\beq
\label{ads5emb}
\rho=r, \quad t=\tau, \quad \phi_1=\Omega t, \quad \zeta=0, \quad \alpha_i=\sigma_i \ . 
\eeq
The intrinsic metric then takes the same  form as in Eq.~\eqref{indmet} now with
\beq
\label{kads}
\bk \equiv |k| =\sqrt{R_0^2-\Omega^2L^2} \ , 
\eeq
where the redshift factor is given by  $ R_0=\sqrt{1+\frac{r^2}{L^2}}$.  Also in this case we find two branches of solutions of the EOMs
\beq
\bar{\Omega}_-=\frac{1}{L} \spa \bar{\Omega}_+=\frac{ \sqrt{9 L^2+8 r^2}}{ 3 } \spa ( 0\le r<\infty ) \ . 
\eeq
The corresponding energy and angular momentum are
\beq
\begin{array}{c}
E_-=T_{\rm D3}\Omega_{(3)}r^2L \spa J_-=T_{\rm D3}\Omega_{(3)}r^2L^2 \ , 
\\[3mm]
E_+=T_{\rm D3}\Omega_{(3)}r^2L^{-1}(3L^2+2r^2) \spa J_+=T_{\rm D3}\Omega_{(3)}r^2L\sqrt{9L^2+8r^2} \ . 
\end{array}
\eeq
In this case the angular momentum is not bounded from above. Moreover, analysis of the off-shell Hamiltonian
shows that the lower branch is $1/2$ BPS and stable while the upper branch is not BPS and unstable.

\subsection{Finite temperature solution}

We now examine the heated up version of the giant graviton moving on $\ads_5$ following the prescription employed in Secs.~\ref{sec:GGthermal} and \ref{sec:GGstab}. Using the embedding \eqref{ads5emb} and the $\ads_5$ background \eqref{adstimess}-\eqref{ads5gauge} we find the EOM 
\begin{equation}
\label{ads5EOM}
\frac{r^2}{L^2}(1-\mathcal{R}_1(\alpha)) + 
3\bk^2 +4 \bk  \frac{r}{L} \mathcal{R}_2(\alpha) =0 \ , 
\end{equation}
which has the following solutions
\beq
\label{Ompmads}
\Omega_\pm=\frac{\sqrt{9L^2+8(1+\Delta_{\pm}(\alpha))r^2}}{3L^2} \ , 
\eeq
with $\Delta_{\pm}(\alpha)$ given in \eqref{delta_pm}. The charge quantization takes the form  \eqref{charge_con} with $\bk$ given by \eqref{kads}. Note that in this case the lower bound on $\alpha$ is given by $\bar{\alpha}$ with $\cosh^2 \bar{\alpha} = 3/2$, which implies that the upper bound on the temperature is $T \leq T_{\rm static}$. This results from the fact that in this case there is no geometric upper bound on $\bk$. Analogously to the $S^5$ case we find 
\begin{equation}
\label{rpma}
\hat{r}_{\pm} (\hat T, \mathbf{k})  =   \frac{ 3 \bk }{ \sqrt{ 1 - 8 \Delta_\pm  (\alpha) }}  \ , 
\end{equation}
with $\Omega_\pm$ given by Eq.~\eqref{Opm} and $\hat{r} = r/L$. 

\subsubsection*{Thermodynamic properties} 

In parallel with \eqref{enes} the energy $E$, angular momentum $J$ and entropy $S$ for this case
are computed to be
\begin{equation}
\label{enesads}
E(r)=   \Omega_{(3)}r^3 \left(R_0^2\frac{\epsilon(r)}{\bk}-\frac{Qr}{L}\right)
\spa 
J(r)= \Omega_{(3)}r^3L^2\frac{\Omega \epsilon(r)}{\bk} 
\spa
S(r)= \Omega_{(3)} r^3 s(r) \ , 
\end{equation}
with $\epsilon (r)$, $s(r)$ given in \eqref{epsilons}. 
The resulting on-shell Gibbs free energy is 
$F =   -  \Omega_{(3)}  (r^3 \bk P + Q r^4/L)  $, and by varying this for constant $T$, $\Omega$ and $Q$ one can obtain the corresponding EOMs \eqref{ads5EOM}. 
Using the definitions of Eq.~\eqref{EJSdef}, we compute the following thermodynamic quantities
\begin{equation}
\mathbf{E}_\pm  (\hat T,\bk) = \frac{ \mathbf{J}_\pm  (1 +  \hat r_\pm^2)  }{\hat \Omega_\pm } - \hat r_\pm^4
\spa 
\mathbf{J}_\pm  (\hat T,\bk) =   
 \frac{27}{16 \sqrt{5}}  \frac{ \bk^3  \hat r_\pm^3  \hat \Omega_\pm  }{\hat T^4} 
 \frac{5+4\sinh^2 \alpha}{\cosh^4 \alpha} \ , 
\end{equation}
\begin{equation}
\label{Fredads}
\mathbf{S}_\pm    (\hat T,\bk)  = \frac{27}{4 \sqrt{5}}  \frac{   \bk^5  \hat r_\pm^3 }{\hat T^5  \cosh^4 \alpha } \spa \mathbf{F}_\pm  (\hat T,\bk)  = 
\frac{27}{16 \sqrt{5}} \frac{ \bk^5  \hat r_\pm^3  }{\hat T^4} 
 \frac{1+4\sinh^2 \alpha}{\cosh^4 \alpha}  - \hat r_\pm^4 \ , 
\end{equation}
where $\hat{\Omega} = \Omega L$. 

\vskip .7cm
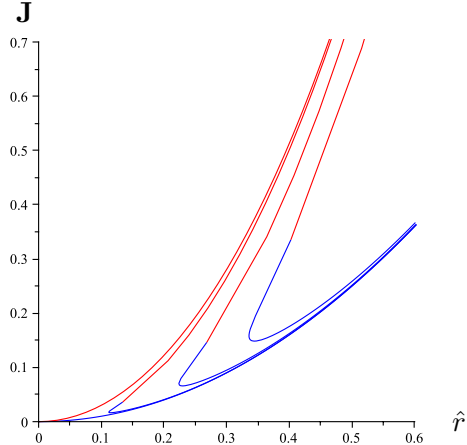
\begin{figure}[!ht]
\centerline{\includegraphics[scale=0.3]{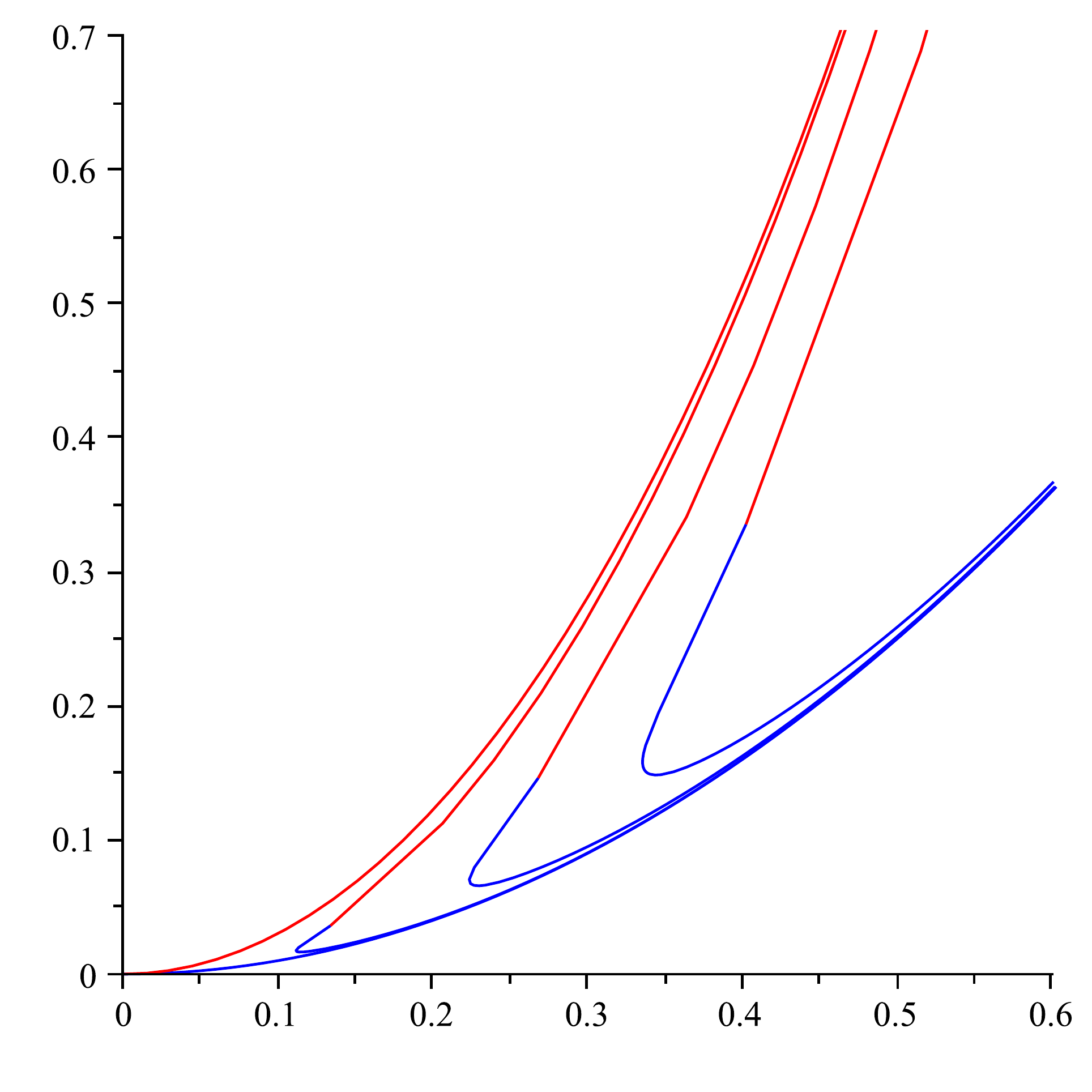}}
\begin{picture}(0,0)(0,0)
\put(310,30){ $\hat r $ }
\put(145,185){ $ \mathbf{J}  $}
\end{picture}		
\vskip -.5cm
	\caption{$\mathbf{J}$   versus  $\hat r $ for the 
	 two solution branches of thermal giant gravitons on $\ads_5$ for $\hat T=0, 0.1,0.2$ and $ 0.3$.} 
	\label{adsfig1}
\end{figure}

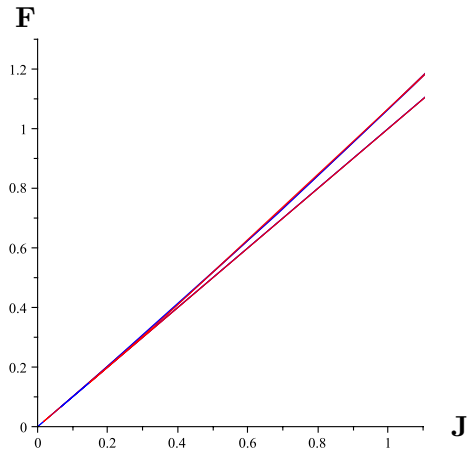
\begin{figure}[!ht]
\centerline{\includegraphics[scale=0.3]{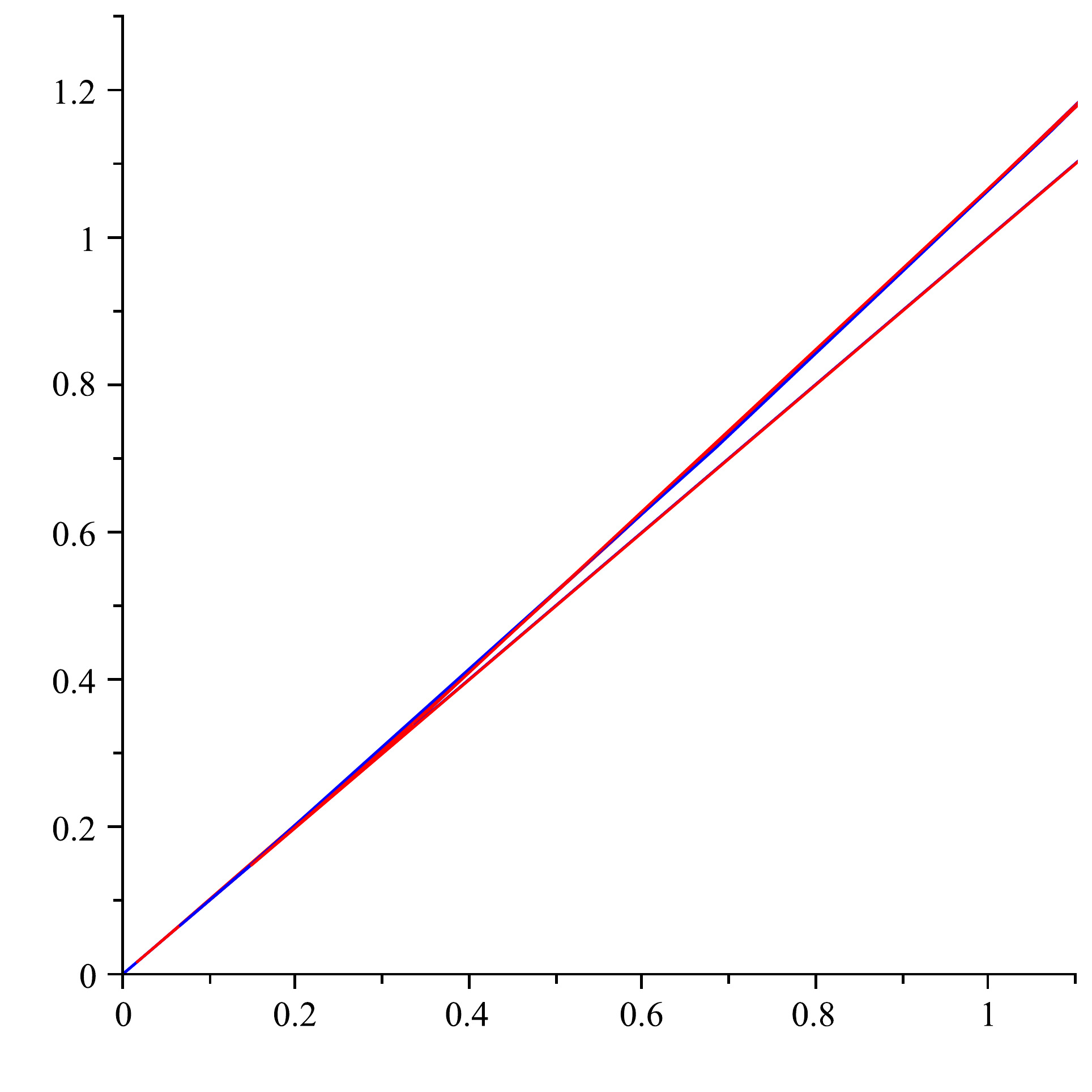}}
\begin{picture}(0,0)(0,0)
\put(310,30){ $\mathbf{J}$ }
\put(145,185){ $ \mathbf{F} $}
\end{picture}		
\vskip -.5cm
	\caption{$\mathbf{F}$   versus  $\mathbf{J}$ for the 
	 two solution branches of thermal giant gravitons on $\ads_5$ for $\hat T=0, 0.1,0.2$ and $ 0.3$. Note that the curves are very close to each other for different $\hat T$.} 
	\label{adsfig2}
\end{figure}

We have plotted the angular momentum $\mathbf{J}$ as function of the $S^3$ radius $\hat{r}$ in Fig.~\ref{adsfig1} as well as the free energy $\mathbf{F}$ as function of $\mathbf{J}$ in Fig.~\ref{adsfig2}. It is clear from Fig.~\ref{adsfig1} that for $T > 0$ the angular momentum is bounded from below as $\mathbf{J} \geq \mathbf{J}_{\rm min} > 0$ as in the case of the thermal giant graviton moving on $S^5$. Correspondingly, the $S^3$ radius $\hat{r}$ is also bounded from below. Instead, there is no upper bound on $\mathbf{J}$ as in the extremal case. 

\vskip .7cm

For $\mathbf{J} > \mathbf{J}_{\rm min}$ there are two available solutions.
From Fig.~\ref{adsfig2} we see that the stable solutions belong to the lower branch but the part of the lower branch between $\hat{r}_{J_{\rm min}}$ and $\hat{r}_{\rm min}$ is unstable. All solutions in the upper branch are unstable, as in the extremal case\footnote{An off-shell analysis in the same spirit of Sec.~\ref{sec:GGstab} using the method described in App.~\ref{app:points} can be carried out giving further evidence for this.}.

The analysis of the validity of the probe approximation works in the same way as for the $S^5$ thermal giant graviton. For small giant gravitons with $r/L$ small we find the condition $r_s \ll r$, or equivalently, $N_{\rm D3}/N \ll r^4 /L^4$. Instead when $r/L$ is not small the condition is $r_s \ll L$ which corresponds to $N_{\rm D3} \ll N$. 
Note that the requirement $T \leq T_{\rm HP}$ implies that $T \ll T_{\rm static}$.

\subsubsection*{Low temperature expansion} 

Focussing on the stable branch of the solution we find that for low temperatures
\begin{equation}
\label{adslowT}
\begin{array}{c} \ds
\Omega_-(\hat T, r) = \frac{1}{L}  + {\cal{O}} (\hat T^8) \ , 
\\[3mm] \ds
\mathbf{J}_- (\hat T,\hat r) = \hat r^2 + \frac{\sqrt{5}}{9 \hat r^2} \hat T^4 + {\cal{O}} (\hat T^8) \ , 
\\[3mm] \ds
\mathbf{F} _-(\hat T,\hat r) = \hat r^2 + \frac{\sqrt{5}}{27}  \frac{3 - \hat r^2}{\hat r^2} \hat T^4+ {\cal{O}} (\hat T^8) \ , 
\end{array}
\end{equation}
Inverting the second equation and plugging into the third equation of \eqref{adslowT} we find exactly the same result for $\mathbf{F}_- (\hat T, \mathbf{J})$  as given in \eqref{FlowerlowT}.
It is interesting that this leading thermal correction to the free energy in the low temperature limit is thus universal for both the lower branches of the thermal giant gravitons on $S^5$ and $\ads_5$.

\section{Discussion and outlook \label{sec:outlook}}

In this paper we constructed and studied thermal giant gravitons. Thermal giant gravitons result from heating up the giant gravitons and the background space-time they move in. We focussed on giant gravitons that in the extremal case are $1/2$ BPS 
with  $ 0 < J \leq N$ (along with a companion non-BPS branch of solutions with $N < J \leq 9N/8$)
obtained from D3-branes wrapped on an $S^3$ and moving in the $\ads_5\times S^5$ background. The thermal giant gravitons are described using black D3-brane probes as described by the blackfold approach 
\cite{Emparan:2007wm,Emparan:2009cs,Emparan:2009at}.

Using the AdS/CFT correspondence our thermal giant graviton solution is expected to correspond to a thermal state in the dual gauge theory. It would be highly interesting to find a description of this thermal state in the gauge theory and compare its properties to those of the thermal giant graviton. In particular, it would be important to compute the free energy correction on the gauge theory side that corresponds to our result Eq.~\eqref{Fresult} for the thermal giant graviton.

An important question that arises from our work is what happens to the system of the thermal giant graviton in $\ads_5\times S^5$ when the temperature is heated up beyond the Hawking-Page temperature $T_{\rm HP}$. For $T < T_{\rm HP}$ the background is in the phase where thermal $\ads_5$ times $S^5$ is dominant. Instead for $T > T_{\rm HP}$ the dominant phase is an AdS black hole times $S^5$. This means the giant graviton moves in the latter background for $T > T_{\rm HP}$. Thus, it would be interesting to repeat the analysis of this paper for the AdS black hole times $S^5$ background.  

One can also study what happens if one considers many giant gravitons moving in $\ads_5\times S^5$. If sufficiently many giant gravitons move along the equator of $S^5$ one can describe them as smeared along this circle. However, while for extremal giant gravitons the descriptions of the smeared and single giant gravitons are equivalent, for thermal giant gravitons the horizon topology would change as one increases the number of giant gravitons to the point where the horizons of each of them overlap. Thus, the non-BPS nature of thermal giant gravitons makes it particularly interesting to study the difference between the smeared and non-smeared phases. Moreover, the smeared phase is connected to the superstar \cite{Myers:2001aq} and 
$1/2$ BPS bubbling AdS solutions of LLM \cite{Lin:2004nb}. Indeed, we expect that the smeared thermal giant graviton solution should be a finite temperature version%
\footnote{See Ref.~\cite{Liu:2007xj} for an exploration of bubbling AdS black holes. Note also that a coarse-graining of LLM geometries has been considered  in \cite{Buchel:2004mc} but these
 do not correspond to supergravity backgrounds with horizons. }
 of the $1/2$ BPS bubbling AdS solution.

Another direction to pursue would be to compute higher-order corrections%
\footnote{See e.g. Refs.~\cite{Emparan:2007wm,Caldarelli:2008pz,Camps:2008hb,Emparan:2009vd,Armas:2011uf,Camps:2012hw} for higher-order corrections in the blackfold approach.}
 in the matched asymptotic expansion to our thermal giant gravitons. In this paper we have worked in the probe approximation $N_{\rm D3} \ll N$ to the leading order in the expansion parameter $N_{\rm D3}/N$. Computing the higher-order corrections would reveal information on what happens for larger values of  $N_{\rm D3}/N$. This could possibly also be interesting to examine numerically. 

Finally, we have focussed in this paper solely on heating up $1/2$ BPS giant gravitons. An interesting future direction to pursue would be to heat up giant gravitons with less supersymmetry \cite{Mikhailov:2000ya}. Note that even in the extremal case not much work has been done to find the explicit brane configurations for $1/4$ and $1/8$ BPS giant gravitons.

\section*{Acknowledgments}

We thank Donovan Young and Costas Zoubos 
for useful discussions. TH thanks NBI and NO thanks Nordita for hospitality. 
JA and NO also wish to thank the KITP for hospitality during the program ``Bits, branes and black holes'',
and acknowledge that this research was supported in part by the National Science Foundation under Grant No. NSF PHY11-25915. The work of NO is supported in part by the Danish National Research Foundation project ``Black holes and their role in quantum gravity''. The work of JA was funded by FCT Portugal grant SFRH/BD/45893/2008.


\begin{appendix}

\section{Detailed stability analysis of extremal giant gravitons \label{app:stability} } 

In this appendix  we carefully analyze the stability of the two solution branches of
extremal giant gravitons by introducing a time dependent perturbation in the radial dynamics and solving the linearized EOMs. More precisely, given a giant graviton configuration characterized by a size $r$ andangular velocity $\Omega$ and with its dynamics  governed by the DBI Langrangean \eqref{dbiL}, we consider the following
perturbation around an on-shell configuration
\beq \label{p1}
r=\hat{r}_0\left(1+\varepsilon \thinspace \mathcal{C}_r \thinspace e^{i\omega\tau} \right) \spa \Omega=\Omega_{0}\left(1+\varepsilon \thinspace \mathcal{C}_\Omega \thinspace e^{i\omega\tau}\right) \ . 
\eeq
Here $\hat{r}_{0}$ and $\Omega_{0}$ denote the on-shell values of the configuration we  perturb around and the perturbation parameter $\varepsilon$ satisfies $\varepsilon\ll1$. The factors $\mathcal{C}_r$ and $\mathcal{C}_\Omega$ measure the amplitude of the oscillations. Our aim is to search for solutions with $\omega^2>0$, signifying the stability of the configuration. 

In order to introduce the perturbations \eqref{p1} in Eq.~\eqref{dbiL} we need to evaluate the quantities involved to lowest order in $\varepsilon$. We begin by obtaining the induced metric $\gamma_{ab}$ for this time-dependent embedding in the form
\beq \label{gammat}
\gamma_{ab}d\sigma^{a}d\sigma^{b}=-\bk^2d\tau^2+r^2d\Omega^2_{(3)}+\mathcal{O}(\varepsilon^2) \ , 
\eeq
where, despite of their resemblance to the unperturbed case, $\bk$ and $r$ are now time-dependent quantities. Doing the same exercise for the WZ term of the action we obtain the time dependent Lagrangean
\begin{equation}
L_{\text{DBI}}=-T_{\rm D3}\Omega_{(3)}r^3\left[\bk-\frac{L^2 \dot{r}^2}{2\rho^2\bk}-r\Omega\right] \ , 
\end{equation}
where we have defined $\rho^2\equiv L^2-r^2$. The two Euler equations for the Lagrangean above take the form
\begin{equation}\label{eqom1}
\bk^2+r^2\Omega^2-4r\Omega \bk+\frac{L^2}{\rho^3\bk^2}\left(r\rho^3\Omega\dot{r}\dot{\Omega}
+r\rho\bk^2\ddot{r}-\frac{r^2}{2}\left((2-L^2\Omega^2)r^2-3L^2\bk^2\right)\dot{r}^2\right)=0 \ , 
\end{equation}
which describes the radial dynamics, while angular momentum conservation is encompassed by
\begin{equation}\label{eqom2}
\frac{\text{d}}{\text{d}\tau}\left(
r^4+\frac{\dot{r}^2r^3 L^2 \Omega }{2 \bk^3}+\frac{\rho^2r^3 \Omega }{\bk} 
\right)=0 \ . 
\end{equation}
In the EOMs \eqref{eqom1}, \eqref{eqom2}, we introduce the perturbations \eqref{p1} and we take the on-shell value $\Omega_{0}$ to be that of the lower or upper branch solution in \eqref{Ommext}. 

 Starting with the lower branch, Eqs.~\eqref{eqom1} and \eqref{eqom2} reduce to
\begin{equation}
\left(\rho^2L C_\Omega \hat{r}_0+2\hat{r}_0 C_r \right)\epsilon+\mathcal{O}(\epsilon^2)=0
\spa \left(2\mathcal C_\Omega+\frac{\mathcal C_r L \hat{r}_0 \omega}{\rho}\right)\epsilon+\mathcal{O}(\epsilon^2)=0 \ , 
\end{equation}
and are solved by $\omega^2=4/L^2$. Therefore, since $\omega^2>0$ for all values of $r$, we conclude that the lower branch is always stable. Due to the 1/2-BPS property of the branch, this is expected. 
For upper branch we instead find the set of equations
\begin{equation}
\left(-2 \mathcal C_\Omega \hat{r}_0+\frac{3\mathcal C_r \rho^2}{\Omega_{0}} \right)\epsilon+\mathcal{O}(\epsilon^2)=0 \  . 
\end{equation}
\begin{equation}
\left(18\mathcal C_\Omega-\mathcal C_r \hat{r}_0 \Omega_0(3L^2\omega^2+16 \rho \Omega_{0}^2) \right)\epsilon+\mathcal{O}(\epsilon^2)=0 \ , 
\end{equation}
which are solved for 
\beq
\omega^{2}=\frac{4}{9L^2\rho \hat \Omega^{2}_{+}}(4r^2-3L^2) \ . 
\eeq
Hence we conclude, as stated in Sec.~\ref{sec:extsol}, that the non-BPS upper branch is stable for $r>\sqrt{3}L/2$ and unstable for $r<\sqrt{3}L/2$.


\section{Thermodynamic blackfold action and Smarr relation \label{app:thermo} }

In this appendix we show that the (mechanical)  action \eqref{BFact} is equivalent to the 
thermodynamic action \eqref{actE}.  To this end we first rewrite \eqref{BFact} as 
\begin{equation}
\label{BFact2}
I = \Delta t \int_{\mathcal B_{p}}dV_{(p)}\left[\mathcal{L}_{\text{(bf)}}+\mathcal{L}_{\text{(em)}}\right] \ , 
\end{equation}
where from now on the subscripts ``bf" and ``em"
refer to the blackfold and external field respectively. 
In \eqref{BFact2} we have factored out the integration over the (Killing) time $t$. This produces a redshift factor which must be included in the Lagrangian densities, e.g. 
$\mathcal{L}_{\text{(bf)}}=\gamma_\perp^{-1}R_0P$ where $\gamma_\perp$ is defined in \eqref{gamma}.  From the conserved quantities derived in \eqref{Mres2}  we also introduce the Hamiltonian and angular momentum densities  
\begin{equation}
\label{eqexp}
\mathcal{H}=\mathcal{H}_{\text{(bf)}}+\mathcal{H}_{\text{(em)}}
=\gamma_\perp^{-1}\left(T^{\mu\nu}_{(\text{bf})}+\mathcal{V}^{\mu\nu}_{(\text{em})}\right)n_\mu\xi_{\nu} \spa
\mathcal{J}=\mathcal J_{\text{(bf)}}+\mathcal J_{\text{(em)}}=\gamma_\perp^{-1} \left(T^{\mu\nu}_{(\text{bf})}+\mathcal{V}^{\mu\nu}_{(\text{em})}\right)n_\mu\chi_{\nu} \ , 
\end{equation}
where $T_{(\text{bf})}^{\mu\nu}$ is the blackfold stress tensor which encapsulates the 
gravitational and electromagnetic self-energy/momentum and    $\mathcal{V}^{\mu\nu}_{(\text{em})} $
(see \eqref{Vdef}) is associated with the coupling of the charge current to the external electromagnetic field. 
Notice that the electromagnetic contributions only depend on the embedding degrees of freedom of the blackfold and not on the effective blackfold fluid degrees of freedom.

Now, for the blackfold degrees of freedom we have the relation
\begin{equation}\label{hamileq1}
\mathcal{H}_{\text{(bf)}}+\gamma_\perp^{-1} u^\mu n_\mu Ts=
\Omega \mathcal \thinspace \mathcal{J}_{\text{(bf)}}-\mathcal{L}_{\text{(bf)}} \ , 
\end{equation}
which follows from eq.(2.19) of \cite{Emparan:2011hg} by multiplying with $\gamma_\perp^{-1}$. 
This is the blackfold generalization of the usual relation $H =\dot{\theta}J-L$ in Hamiltonian mechanics  between Hamiltonian and Lagrangian, but now with an extra term contributing to the energy due to the fact that the blackfold has internal thermal degrees of freedom living on it.  However, since the external electromagnetic field does not couple to the thermal degrees of freedom 
living on the blackfold, one has for the electromagnetic part  that  
\begin{equation}\label{hamileq}
\mathcal{H}_{\text{(em)}}=\Omega \thinspace \mathcal J_{\text{(em)}}-\mathcal{L}_{\text{(em)}} \ . 
\end{equation}
We now use \eqref{hamileq1}, \eqref{hamileq} in \eqref{BFact2} along with the expression \eqref{BFentropy} for
the total  entropy $S$ of the blackfold.  If we also rotate to Euclidean time so that $\Delta t \rightarrow \Delta \tau = \beta = 1/T$, we then find that the Euclidean action is given by 
\begin{equation}
I_E =E-\Omega J-TS \ . 
\end{equation}
Here we recall that, as in \cite{Emparan:2011hg} this is the Euclidean action at fixed charge $Q_p$. 
As explained in that reference, it is also possible to go to an ensemble where the charge can vary
by introducing a potential  $\Phi_p$ dual to the charge and performing a Legendre transformation. 

\subsubsection*{Smarr relation} 

Finally, we derive the Smarr formula for blackfolds in external fields. We use the perfect fluid stress tensor
$T_{\mu \nu} = (\epsilon +P ) u_\mu u_\nu + P h_{\mu \nu}$  and the local thermodynamic relations
for charged $p$-branes in $D=n+p+3$ dimensions
\beq
\epsilon+P=\mathcal{T}s \spa \epsilon=-(n+1)P-n\Phi_p Q_p \ . 
\eeq
First, we note that the Smarr relation found previously for blackfolds based on charged $p$-branes
(with zero external field) is easily generalized to the case where $\xi^{\mu}$ is not orthogonal to the worldvolume
$\CB_p$. One finds 
\beq
\label{Smarr1}
(D-3)E_{\text{(bf)}}-(D-2)\left(\Omega J_{\text{(bf)}}+TS\right)-n\Phi_H Q_p=\mathcal{T}_{\text{(bf)}}^{\rm tot} \  , 
\eeq
where
\beq
\Phi_H=\int_{\mathcal{B}_p} dV_{(p)} \gamma_\perp^{-1} R_0 \Phi_p \ , 
\eeq
\beq
\mathcal{T}_{\text{(bf)}}^{\rm tot}=-\int_{\mathcal{B}_p} dV_{(p)}\left(
\gamma_\perp^{-1} R_0 
 {\rm tr}\,T+\gamma_\perp^{-1}T^{\mu\nu}_{\text{(bf)}}\xi_{\mu}n_{\nu}\right) \spa {\rm tr}\,T \equiv \gamma_{ab}T^{ab} \ . 
\eeq
We then add to both sides of \eqref{Smarr1} the term $(D-3)E_{\text{(em)}}-(D-2)\Omega J_{\text{(em)}}$,
yielding the generalized Smarr relation 
\beq \label{Smarr}
(D-3)E-(D-2)\left(\Omega J+TS\right)-n\Phi_H Q_p=\mathcal{T}_{\rm tot} \ , 
\eeq
where
\beq
\mathcal{T}_{\rm tot}=-\int_{\mathcal{B}_p} dV_{(p)}\left(\gamma_\perp^{-1}R_0 {\rm tr}\,T+\gamma_\perp^{-1}(T^{\mu\nu}_{\text{(bf)}}+\mathcal{V}_{\text{(em)}}^{\mu\nu})\xi_{\mu}n_{\nu}+(D-2)\mathcal{L}_{\text{(em)}}\right) \ . 
\eeq
Note that, as expected, the total tension gets modified by the presence of the external field.

\section{Derivation of the thermal giant graviton equation of motion \label{app:EOMthermal} }

In this appendix we provide some details on the derivation of the thermal giant graviton blackfold EOM \eqref{exteq}.
First we note that for a perfect fluid stress tensor, the extrinsic blackfold EOM \eqref{gen_extr_eq} can 
be rewritten as \cite{Emparan:2009at,Emparan:2011hg} 
\begin{equation}
\label{exteq0} 
PK^\mu+s\mathcal{T}\dot{u}^\mu=\mathcal{F}^\mu \ , 
\end{equation}
where $P$ is the pressure, $s$ the entropy density, $\mathcal{T}$ the local temperature, 
$K^\mu$ the extrinsic curvature vector, $\dot u^\mu$ the fluid acceleration 
 and $\mathcal{F}^\mu$ the external force. 

We proceed by computing the various terms. Using  \eqref{Kdef} along with the
the background \eqref{ds5} and  embedding \eqref{D3embed} we compute
 \begin{equation}
K_{\tau\tau}^{\ \ \zeta} =\frac{\Omega^2r\mu_1}{L^2} \spa
K_{11}^{\ \ \zeta}=-\frac{\mu_1\mu_2^2}{rL^2} \spa
K_{22}^{\ \ \zeta}=-\frac{\mu_1\mu_3^2}{rL^2} \spa
K_{33}^{\ \ \zeta}=-\frac{r\mu_1}{L^2} \ , 
\end{equation}
where
\begin{equation}
\mu_1=\sqrt{L^2-r^2} \spa \mu_2=r \sin \theta \spa  \mu_3=r \cos \theta  \ , 
\end{equation}
and we recall that $\zeta$  is the direction parameterizing the fibration of the $S^5$ into $S^3$Õs 
on which the giant graviton of size $r=L \sin \zeta$  is defined. 
This gives for the extrinsic curvature vector 
\begin{equation} \label{meanc}
K^{\zeta}=-\frac{\mu_1}{r} \frac{\Omega^2r^2+3\bk^2}{L^2\bk^2} \ , 
\end{equation}
with the other components of $K^\mu$ equal to zero. Notice that $K^\zeta$ is manifestly negative. 
To compute the extrinsic blackfold force term, we first compute the pull back of the RR field strength \eqref{F5ds5} 
on the blackfold according to  $  F_{\tau 123}{}^{\zeta}=4\Omega  d \Omega_{(5)} /L$. 
Using this together with the blackfold four-current \eqref{D3current} in the right side of \eqref{gen_extr_eq},
 we find that the only non-vanishing component of the  force term is
\begin{equation}
\label{bfforce}
\mathcal F^\zeta=\frac{4\Omega Q \mu_1}{L^2\bk} \ . 
\end{equation}
We see that $F^\zeta/(\Omega Q)$ is manifestly positive. 
Finally we compute the fluid acceleration $\dot{u}^\mu=u^\nu \nabla_\nu u^\mu$. To this end
notice that although the (local boost) vector field $u^a$ is only defined on the 
world volume of the giant graviton, it can be pushed forward to the vector field $u^\mu$ 
on the entire $\mathbb{R}\times S^5$. This means that the acceleration can be computed using $\dot{u}^\mu=\partial^\mu \log \bk $ so that 
\begin{equation}
\label{uacc}
\dot{u}^\zeta=\frac{\Omega^2\mu_1r}{L^2\bk^2} \ , 
\end{equation}
with the rest of the components of $\dot{u}^\mu$ equal to zero. Note that the acceleration $\dot{u}^\zeta$ is manifestly positive.

The extrinsic equation \eqref{exteq0} is thus only non-trivial for $\mu$ in the $\zeta$-direction 
and from the results above we see that since the blackfold pressure $P$ (see \eqref{thermoquantities}) is negative, the left hand side is manifestly positive. The structure is therefore clear: In order for the D3 brane not to collapse under gravity, the electromagnetic
repulsion term (the right hand side) must exactly balance the gravitational pull (the left hand side). 
We  therefore conclude that  $\Omega Q>0$, which means that the solution must always be rotating and charged, as expected. 
Moreover, we see that  the solution space is symmetric under charge conjugation and time inversion. 

The thermal giant graviton blackfold EOM \eqref{exteq} of the text is then obtained from \eqref{exteq0}, by
substituting \eqref{meanc}, \eqref{bfforce}, \eqref{uacc}  and using the pressure $P$ in \eqref{thermoquantities}
along with $\mathcal{T}$ and $s$ in \eqref{TandS}.

\section{Analysis of the two meeting points \label{app:points} } 

In this appendix we perform a careful analysis of the two meetings points in configuration space mentioned in Sec.~\ref{sec:GGstab} corresponding to the maximal and minimal charge parameter giant gravitons. These bifurcation points (see e.g. Ref.~\cite{Arcioni:2004ww}) deserve special attention as they provide key information about the overall stability properties of the thermal giant graviton. Being bifurcation points where two sets of equilibria configurations meet, the stability properties of the physical system can in generally change and therefore special attention is needed. 

According to the stability analysis that was presented in Sec.~\ref{sec:GGstab} for the low temperature regime, all configurations near the maximal size are stable in both branches. However, one can imagine increasing slightly the temperature and moving away from such regime while still being below the Hawking-Page temperature. The analysis of the stability properties of the maximal giant graviton carried out below allow us to conclude that, even after the temperature is further increased, the thermal gravitons are stable. Moreover, part of the lower branch, which contains $\mathbf{J}_{\text{min}}$, is not covered by the low temperature expansion. Since we do not reach $\mathbf{J}_{\text{min}}$ analytically 
in that expansion, we instead analyze here the end point of the lower branch described by the minimal charge parameter giant graviton and show explicitly that a change of stability has occurred. We begin by describing these two different limiting cases and then proceed to study their stability in Sec.~\ref{stab_method}.

\subsection{The maximal giant graviton limit}  \label{max_giant}
In this section we probe the regime in solution space dominated by the dynamics of the maximal giant graviton, which is achieved by performing an expansion of Eqs.~\eqref{rpm}-\eqref{Opm} around $\bk =1$.  The extremal maximal giant graviton is the object for which there is good evidence that its dual operator is a Schur polynomial, hence 
this configuration is expected to be an ideal candidate for a preliminary study on the dual state of giant gravitons at finite temperature. In view of this, we begin by describing the properties of the thermal maximal giant graviton. At the exact point $\bk =1$ (or $r=L$) the angular velocity of the configuration is given by
\beq
\Omega_{\pm}(\phi_0)=\frac{3}{L\sqrt{1-8\Delta_{\pm}(\phi_{0})}} \ , 
\eeq
where $\phi_{0}$ is given in terms of $\delta_0$ is obtained by setting $\bk =1$ in Eq.~\eqref{cosdelta}, \eqref{alphasol}. The conserved charges can be obtained from Eqs.~\eqref{enes}. The total energy and angular momentum then take the form
\beq
\textbf{E}_{\pm}(\phi_0)=\frac{27}{16\sqrt{5}}\frac{\phi_0}{\hat{T}^4}(4+\phi_0) \spa \textbf{J}_{\pm}(\phi_0)=1 \ , 
\eeq
where we see clearly that, as in the extremal case, the angular momentum is independent of the angular velocity and temperature of the configuration. Moreover the total entropy and Helmholtz free energy read
\beq
\textbf{S}_{\pm}(\phi_0)=\frac{27}{16\sqrt{5}}\frac{\phi_0^2}{\hat{T}^5} \spa \textbf{F}_{\text{H}}^{\pm}(\phi_0)=\frac{27}{16\sqrt{5}}\frac{(4-3\phi_0)\phi_0}{\hat{T}^4} \ . 
\eeq

\subsubsection*{The large giant graviton expansion}

We now perform the expansion around $\bk=1$. From Eq.~\eqref{rpm} we find a relation between the values of $\hat r$ and the values of $\bk$ of the form 
\beq
\hat{r}_\pm=1+\frac{\bk-1}{\hat{\Omega}_{\pm}(\phi_0)} \ . 
\eeq
Parametrizing the expansion in terms of $\hat{r}$ we invert the above relation in order to find:
\beq
(\bk_{\pm}-1)=\hat{\Omega}^{2}_{\pm}(\phi_0)(\hat{r}-1) \ . 
\eeq
Since the expansion is only valid for values of $\bk \sim 1$ we must require $L^{2}\Omega^{2}_{\pm}(\phi_0)(\hat{r}-1)\ll 1$. Defining $d\hat{r}\equiv \hat{r}-1$ and using Eq.~\eqref{Opm} to perform the same expansion yields
\beq
\Omega_{\pm}=\Omega_{\pm}(\phi_0)\left(1+\frac{\hat{\Omega}^{4}_{\pm}(\phi_0)}{9}g(\phi_0)d\hat{r}\right) \ , 
\eeq
where we have defined the function $g(\phi_0)$ through the expression:
\beq
g(\phi_0)=-8(1+\Delta_{\pm}(\phi_0))(-1+\Delta_{\pm}(\phi_0))+36\partial_{\bk}\Delta_{\pm}(\phi_0) \ . 
\eeq
Similarly, the charge parameter $\phi$ given in Eq.~\eqref{alphasol} is expanded to
\beq
\phi=\phi_0\left(1-\hat{\Omega}_{\pm}(\phi_0)\phi_0 f(\phi_0)d\hat{r}\right) \ , 
\eeq
where we have defined the function $f(\phi_0)$ as
\beq
f(\phi_0)=2\left(2\sqrt{3}\text{cos}(\frac{2}{3}\delta_{0})+\sqrt{3}\text{cos}(\frac{4}{3}\delta_0)+8\text{cos}(\frac{\delta_0}{3})\text{sin}^3(\frac{\delta_0}{3})\right)\text{sec}(\delta_0)\text{csc}(\delta_0) \ . 
\eeq
The physical properties can be easily obtained from expressions \eqref{enes} and read
\beq
\textbf{E}_{\pm}=\textbf{E}_{\pm}(\phi_0)\left(1+\left[3\left(1+\hat{\Omega}^2_{\pm}(\phi_0)\right)-\hat{\Omega}^2_{\pm}(\phi_0)f(\phi_0)\frac{\phi_0-3}{4+\phi_0}\right]d\hat{r}\right) \ , 
\eeq
\beq
\textbf{J}_{\pm}=\textbf{J}_{\pm}(\phi_0)+\left(4-2\hat{\Omega}_{\pm}(\phi_0)\phi_0(4+\phi_0)\right)d\hat{r} \ , 
\eeq
\beq
\textbf{S}_{\pm}=\textbf{S}_{\pm}(\phi_0)\left(1+\left[3\left(1+\hat{\Omega}^2_{\pm}(\phi_0)\right)+2\hat{\Omega}^2_{\pm}(\phi_0)(1-f(\phi_0))\right]d\hat{r}\right) \ , 
\eeq
and finally the Helmholtz free energy
\beq
\! \! \! \! \textbf{F}^{\pm}_{\text{H}}=\textbf{F}^{\pm}_{\text{H}}(\phi_0)\left(1+\left[\frac{-12+9\phi_0+\hat{\Omega}^2_{\pm}(\phi_0)(-12+17\phi_0-54(2-3\phi_0)f(\phi_0))}{4-3\phi_0}\right]d\hat{r}\right) \ . 
\eeq

\subsection{The minimal charge parameter limit} \label{min_giant}

The minimal charge parameter limit is the point in configuration space where the two branches connect smoothly and can be seen as the point particle analog at finite temperature. Furthermore, it is a limit which describes thermal giant graviton configurations at any temperature $T$, including very low temperatures, but it is not captured by the low temperature limit of Sec.~\ref{sec:GGstab} because $\hat{r}\sim \hat{T}$. At the exact meeting point the thermal giant graviton is characterized by a specific value of the charge parameter $\tilde\alpha$ which implies a characteristic size $\hat{\tilde{r}}$ given by \eqref{rtilde} and corresponding angular velocity
\beq
\hat{\tilde{\Omega}}=\sqrt{1+\frac{4}{5}\hat{T}^2} \ . 
\eeq
The physical properties are then easily obtained from Eqs.~\eqref{enes} and read:
\beq
\tilde{\textbf{E}}=\frac{18}{5}\frac{\hat{T}^2}{\hat{\tilde{\Omega}}^3} \spa
\tilde{\textbf{J}}=2\hat{\tilde r}^2-\hat{\tilde r}^4 \spa \tilde{\textbf{S}}=\frac{4}{9}\hat{\tilde r}^4\hat{\tilde \Omega}
\spa 
\tilde{\textbf{F}}_{\text{H}}=\frac{18}{25}\frac{5-2\hat{T}^2}{\hat{\tilde \Omega}^2}\hat{T}^2 \ . 
\eeq

\subsubsection*{The minimum charge parameter expansion} 

We now obtain an effective description of the physics near this limit by expanding the physical properties of these configurations near $\bk =\hat{T}$. We begin by expanding the charge parameter parametrized by $\phi$ using Eq.~\eqref{alphasol}, which to leading order yields
\beq
\phi=\frac{4}{9}-\mathcal{C}^2\left(\frac{\bk}{\hat{T}}-1\right) \ , 
\eeq
where $\mathcal{C}$ is a numerical constant ($\sim \sqrt{3}$) which can be calculated exactly. Using this form of the charge parameter we expand the size $r$ using Eq.~\eqref{rpm}, which leads to
\beq
r_{\pm}=\tilde r\left(1+\frac{3\sqrt{5}\mathcal{C}}{\sqrt{2}}\frac{1-\hat{T}^2}{5+4\hat{T}^2}\sqrt{\frac{\bk}{\hat{T}}-1}\right) \ . 
\eeq
Thus we invert the equation above to find the relation
\beq
\left(\frac{\bk_\pm}{\hat{T}}-1\right)^{\frac{1}{2}}=\pm\frac{\sqrt{10}}{3\mathcal{C}}d\hat{\tilde r} \ , 
\eeq
where we have defined the expansion parameter
\beq
d\hat{\tilde r}=\frac{1}{1-\hat{\tilde r}}\left(1-\frac{r}{\tilde r}\right) \ . 
\eeq
Therefore for this approximation to be valid we need to require $d\hat{\tilde r}\le1$. Using the above definition we can rewrite the correction to the angular velocity to leading order in the form
\beq
\Omega_{\pm}=\tilde{\Omega}\left(1\pm\frac{9\hat{T}^2}{5\hat{\tilde \Omega}^2}d\hat{\tilde r}\right) \ . 
\eeq
It is then straightforward to write down all the physical properties using Eqs.~\eqref{enes}
\beq
\textbf{E}=\tilde{\textbf{E}}\left(1\pm\frac{3(\hat{T}^2-1)}{\hat{\tilde \Omega}^2}d\hat{\tilde r}\right) \spa \textbf{J}=\tilde{\textbf{J}}\left(1\pm\frac{6(5+2\hat{T}^4-7\hat{T}^2)}{\hat{\tilde \Omega}^2(10-\hat{T}^2)}d\hat{\tilde r}\right) \ , 
\eeq
\beq
\textbf{S}=\tilde{\textbf{S}}\left(1\pm\frac{6(1-\hat{T}^2)}{\hat{\tilde \Omega}^2}d\hat{\tilde r}\right)
\spa \textbf{F}_{\text{H}}=\tilde{\textbf{F}}_{\text{H}}\left(1\pm\frac{3(5+2\hat{T}^4-7\hat{T}^2)}{\hat{\tilde \Omega}^2(5-2\hat{T}^2)}d\hat{\tilde r}\right) \ . 
\eeq


\subsection{Stability properties in the various limits} \label{stab_method}
In this section we examine the stability properties of the various limits of the thermal giant graviton taken in the previous sections. To this aim we consider the localized giant graviton to be in thermodynamical equilibrium with the surroundings at temperature $T$. Moreover, since the total angular momentum $J$ is conserved, the relevant variables for describing the thermodynamic ensemble are the size of the giant graviton $r$, the temperature $T$, the angular momentum $J$ and the (conserved) total charge $Q=T_{\rm D3}N_{\rm D3}$. The stable solutions to the blackfold EOMs are then characterized by the paths in configuration space for which the Helmholtz free energy $F_{\text{H}}=E-TS$ is minimized for $T$, $J$ and $Q$ held fixed. In other words, the stable solutions are determined by the requirements
\begin{equation}\label{derivatives}
(F_{\text{H}})_{(1)}\equiv \frac{\partial F_\text{H}}{\partial r}\Big|_{T,J,Q}=0 \quad \text{and} \quad (F_{\text{H}})_{(2)}\equiv \frac{\partial^2F_\text{H}}{\partial r^2}\Big|_{T,J,Q}>0 \ . 
\end{equation}
The first of these equations is equivalent to the EOM \eqref{exteq} and was examined in Sec.~\ref{sec:GGthermal}. The formulae \eqref{enes} for the conserved quantities allows us to obtain the free energy of a (in general off-shell)
given thermodynamical configuration. However, notice that the free energy will be parameterized in terms of the angular velocity $\Omega$ and charge parameter $\phi$. Determining the 
derivatives \eqref{derivatives} is straight forward. For a given $r$, $T$ and $Q$, let $\Omega=\Omega(r;T,Q)$ and $\phi=\phi(r;T,Q)$ denote the corresponding on-shell values. Now consider a small variation of the configuration $r\to r+dr$, $\Omega\to \Omega(r;T,Q)+\Omega_{(1)}\text{d}r+\Omega_{(2)}\text{d}r^2$,  $\phi\to \phi(r;T,Q)+\phi_{(1)}\text{d}r+\phi_{(2)}\text{d}r^2$ so that $J$, $T$ and $Q$ are kept constant up to $\mathcal{O}(\text{d}r^2)$. This gives us four equations (using equations \eqref{enes} and \eqref{charge_con} for each order) which allows us to determine the four parameters $\Omega_{(1)}, \Omega_{(2)}, \phi_{(1)}, \phi_{(2)}$. Since we are perturbing around an on-shell 
configuration $(F_{\text{H}})_{(1)}=0$, so that the overall change in $F_\text{H}$ is 
\begin{equation}
F_\text{H}\to F_\text{H}+(F_{\text{H}})_{(2)}\text{d}r^2 \ . 
\end{equation}
Inspecting the sign of $(F_{\text{H}})_{(2)}$ allows us to determine the stability of the given solution to the blackfold EOM. This method was used in Sec.~\ref{sec:lowTlimit} to probe the stability properties in the low temperature regime. We now make further use of this method by applying it to the two cases presented in the previous sections and state the results for the second order change in the free energy in the various limits.

\subsection*{Maximal giant graviton limit}
Close to maximality we make use of the expansion given in Sec.~\ref{max_giant}. We find for the second order change in the free energy
\beq
\! \! \! (\textbf{F}_{\text{H}})_{(2)}^{\pm}=\frac{4}{\hat{\rho}^2}\frac{4-4\phi_0-\hat{\Omega}_{\pm}(\phi_0)\sqrt{1-\phi_0}(4+\phi_0)+\hat{\Omega}^2_{\pm}(\phi_0)(1+\frac{\phi_0}{4})^2}{\sqrt{1-\phi_0}(4+\phi_0)} + H(\phi_0,\hat{r})d\hat{r} \ , 
\eeq
where $H(\phi_0,\hat{r})$ is some intricate function of $\phi_0$ and $\hat{r}$. An explicit computational check can be performed in order to conclude that $(\textbf{F}_{\text{H}})_{(2)}^{\pm}>0$ for all values of $r$ except at $\hat{T}=1$ (see below). Hence, near maximality, all giant graviton configurations are stable and so remain for all temperatures.

\subsection*{Minimal charge parameter limit}
At the bifurcation point in configuration space where the two branches meet smoothly, the expansion given in 
Sec.~\ref{min_giant} is the required tool to study the stability properties of these configurations. Using the same method as for the previous case we find the second order change in the free energy:
\beq
(\textbf{F}_{\text{H}})_{(2)}^{\pm}=(\tilde{\textbf{F}}_{\text{H}})_{(2)}\left(1\pm\frac{5\sqrt{5}\hat{\tilde \Omega}}{6(5-2\hat{T}^2)}\hat{T}^2d\hat{\tilde r}\right) \spa (\tilde{\textbf{F}}_{\text{H}})_{(2)}=-\frac{12\hat{\tilde \Omega}^2}{5-2\hat{T}^2}{\tilde{\mathbf{F}}_\text{H}} \ . 
\eeq
Careful inspection of the sign of $(\textbf{F}_{\text{H}})_{(2)}^{\pm}$ leads us to conclude that all giant graviton configurations near the bifurcation point are unstable, i.e., $(\textbf{F}_{\text{H}})_{(2)}^{\pm}<0$. This provides further evidence for the change of stability properties occurring at $\textbf{J}_{\text{min}}$ and $\textbf{J}_{\text{max}}$ as explained in Sec.~\ref{sec:GGstab}.


 \end{appendix}

\addcontentsline{toc}{section}{References}

\providecommand{\href}[2]{#2}\begingroup\raggedright\endgroup

\end{document}